\documentclass[11pt]{article}
\pdfoutput=1 

\usepackage{jheppub} 

\usepackage{jheppub} 
\usepackage[T1]{fontenc} 
\usepackage{physics}
\usepackage{float} 
\usepackage{subcaption}
\usepackage{xcolor}

\title{\boldmath Complex CFTs: Holography and Interfaces}

\author{Luis Camargo-Carlos and}
\author{Michael Gutperle}

\affiliation{
Mani L. Bhaumik Institute for Theoretical Physics, Department of Physics and Astronomy,\\ University of California, Los Angeles, CA 90095, USA
}

\emailAdd{lcamargo2@g.ucla.edu}
\emailAdd{gutperle@ucla.edu}

\abstract{We investigate complex conformal field theories and  interfaces between such theories, using methods of holography and two dimensional CFT. We use holography to prove the Im-flip property of complex conjugate CFTs. We construct a  complex Janus solution and calculate some holographic observables. Complex RG-flow interfaces are  obtained numerically. On the CFT side, complex interfaces are constructed for free boson CFTs with and without background charge.  Boundary states are constructed, and transmission and reflection coefficients are calculated in  both cases.}

\begin{document} 
\maketitle
\flushbottom 


\section{Introduction}

Complex conformal field theories are not unitary but can nevertheless have uses in physics. In  \cite{Gorbenko:2018ncu, Gorbenko:2018dtm} it was argued that complex CFTs are useful to describe the phenomenon of walking \cite{Kaplan:2009kr} which can be illustrated by the following beta function for the renormalization group flow of a single coupling $\lambda$:
 
\begin{align}\label{betafirst}
    {d\lambda \over d t} =\beta(\lambda) = - x-\lambda^2 + o(\lambda^3)
\end{align}
For small values of the coupling the behavior of the flow depends on the  sign of $x$. For $x<0$ there are two real fixed points, one IR and UV attractive respectively.
For small positive $x$ there are no real fixed points and the flow of $\lambda$ near the origin is described as "walking", i.e. the coupling stays there for a long RG-flow time.  If one allows the coupling to move into the complex plane for the $x>0$ case there are two complex conjugate fixed points $\lambda=\pm i \sqrt{x}$. It was argued in \cite{Gorbenko:2018ncu} that the complex CFTs control the real flow and are also interesting subject to study in their own right.

An  interesting  example in two dimensions is the $Q$-state Potts model. For integer  $Q<4$, the theory at its critical point is a minimal model, whereas for $Q=4$ it is described by an orbifold CFT of a single free boson.  For the Potts model with $Q>4$ the phase transition at the critical point becomes first order and conformality is lost. However,  for $Q$ larger but close to $Q=4$ the theory exhibits walking behavior. In \cite{Gorbenko:2018dtm} the two CFTs associated with the complex fixed points were studied. The realization of these complex CFTs in lattice models was studied in \cite{Jacobsen:2024jel,Tang:2024blm,Ma:2018euv}. Similar phenomena also occur in two dimensional $O(n)$ models with $n>2$ \cite{Haldar:2023ukr} and higher dimensional gauge theories \cite{Giombi:2019upv,Benini:2019dfy} and deconfined quantum critical points \cite{Wang:2017txt,Senthil:2023vqd}.

The main goal of the present paper is to study aspects of walking CFTs and the associated complex CFTs from a holographic point of view. We review the Im-flip relation for complex conjugate CFTs argued for in \cite{Gorbenko:2018ncu} and present a holographic proof of the relation. Furthermore  we consider holographic conformal interfaces between theories which are complex conjugate, investigating how observables such as the central charge, defect entropy and reflection and transmission coefficients behave for such interfaces. We also present CFT constructions of conjugate interfaces for noncompact  free bosons with and without background charge.

\subsection{Im-flip relation}

In this section we review an argument presented in \cite{Gorbenko:2018ncu} which provides a relation between the imaginary part of scaling dimensions and OPE coefficients for the complex conjugate CFTs governing the walking behavior.   The two complex fixed points define two CFTs  $\mathcal{C}$ and $\bar{\mathcal{C}}$. It follows from expanding  the beta function (\ref{betafirst}) that the theory $\mathcal{C}$, has an almost marginal operator $\mathcal{O} $ which has scaling dimension $\Delta_{\mathcal{O}} = 2 - i\epsilon$. Note we specialize to  two dimensional CFTs here to make contact with later sections, but the result straightforwardly generalizes to other dimensions.
The theory $\mathcal{C}$ is deformed by this operator
\begin{align}
S_{\mathcal{C}}\to S_{\mathcal{C}}+g\int d^dx\mathcal{O}(x)
\end{align}
Using standard results of conformal perturbation theory the beta function for the coupling $g$ becomes 
\begin{align}
    \beta(g) = -i\epsilon g + \frac{\pi}{2}C_{\mathcal{O}\mathcal{O}\mathcal{O}}g^2
\end{align}
Where $C_{\mathcal{O}\mathcal{O}\mathcal{O}}$ is the  three point function of $\mathcal{O}$.
The fixed points of this equation are then at $g = 0$, which is the unperturbed CFT $\mathcal{C}$ and a second fixed point $g_{FP} = \frac{2i\epsilon}{\pi C_{\mathcal{O}\mathcal{O}\mathcal{O}}}$. 
We introduce a second generic operator which we denote $\psi$, the relevant data is the conformal dimensions $\Delta_\psi$ and the $\mathcal{O} \psi \psi$ OPE coefficient $C_{\mathcal{O} \psi \psi}$.
The claim of \cite{Gorbenko:2018ncu}  is that the fixed point $g_{FP}$ corresponds exactly to $\bar{\mathcal{C}}$. 
It follows from this statement  that the following identity must be true:

\begin{equation}\label{imfliprel}
    \frac{Im(\Delta_{\mathcal{O}})}{C_{\mathcal{O}\mathcal{O}\mathcal{O}}} = \frac{Im(\Delta_{\psi})}{C_{\mathcal{O}\psi\psi}}\\
\end{equation}

Note the OPE coefficient $C_{\mathcal{O}\mathcal{O}\mathcal{O}}$ and $C_{\mathcal{O}\psi\psi}$ are in general complex, however at $g=0$ they are real and they only pick up an imaginary part at higher order in conformal perturbation theory which we drop since we want proof (\ref{imfliprel}) to lowest nontrivial order. 
 Note that the identity above relies on the identification of the complex conjugate theory $\bar{\mathcal{C}}$ with the theory at the nontrivial fixed point $g_{FP}$. Without such an identification, there would be a proliferation of other complex fixed point CFTs which is argued to be unphysical \cite{Gorbenko:2018ncu}.

First, note that the identification of $\bar{\mathcal{C}}$ with the theory at $g_{FP}$ implies that for any operator $\phi$, 
\begin{equation}
    \Delta_{\psi}(0) = \overline{\Delta_{\psi}(g_{FP}}) \implies Im(\Delta_{\psi}(0)) = Im(\overline{\Delta_{\psi}(g_{FP})} = - Im(\Delta_{\psi}(g_{FP}))\\
\end{equation}
From this one can calculate
\begin{align}
        Im(\Delta_{\psi}(0)) &= -Im(\Delta_{\psi}(0)) - Im(\pi C_{\mathcal{O}\psi\psi}g_{FP})\nonumber\\
        2Im(\Delta_{\psi}(0))&=-2Im\left(\frac{i\epsilon C_{\mathcal{O}\psi\psi}}{C_{\mathcal{O}\mathcal{O}\mathcal{O}}}
        \right)
    \end{align}
To first order in $\epsilon$, the above simplifies to 
\begin{equation}
    \begin{split}
        \frac{Im(\Delta_{\psi}(0))}{C_{\mathcal{O}\psi\psi}} &= -\frac{\epsilon }{C_{\mathcal{O}\mathcal{O}\mathcal{O}}}
    \end{split}
\end{equation}
Note that $-\epsilon= Im(\Delta_{\mathcal{O}}(0))$, so at  $g = 0$ we have the desired identity (\ref{imfliprel}). In the following we generalize the argument along the flow, where a generic operator $\phi$ picks up an anomalous dimension
\begin{equation}
    \gamma_g = \pi C_{\mathcal{O}\psi\psi} \; g+\mathcal{O}(g^2)\\
\end{equation}

First, consider the case when  $g$ is purely imaginary. Then, up to first order in $\epsilon$, the OPE coefficient will be taken to be real. So $Im\gamma_g$ will be the same for all operators up to an OPE coefficient. Second consider the case when  $g$ has a real part. This will also be small, so again $Im\gamma_g$ will be the same for all operators up to the OPE coefficient.  Then, at some general point along the flow $g'$, 
\begin{align}
        \frac{Im(\Delta_{\psi}(g))}{C_{\mathcal{O}\psi\psi}} &= \frac{Im(\Delta_{\psi}(0)) + \pi C_{\mathcal{O}\psi\psi}Im(g')}{C_{\mathcal{O}\psi\psi}}\nonumber \\
        &=\frac{Im(\Delta_{\psi}(0)) }{C_{\mathcal{O}\psi\psi}} +\pi Im(g')
\end{align}

The additional term is the same for all operators, so the statement holds for a  general value along the flow. It is worth noting that the above argument relied on two ideas: First,  the imaginary parts are small in OPE coefficients and anomalous dimensions, and second, There is a complex conjugate theory $\bar{\mathcal{C}}$ that is exactly the theory at the nontrivial zero of $\beta(g)$. The small imaginary part is a necessary condition for a real flow to exhibit walking behavior. For larger imaginary parts, the conformal perturbation theory will break down, and it is unclear if walking will persist. The assumption  of  nonproliferation of complex fixed points is physically very reasonable, but there is no independent proof at this point of which we are aware.

\medskip

{\bf Note:} While finalizing this manuscript a paper \cite{Furuta:2026eua,Tang:2026ccf} appeared which has some overlap with section \ref{complexcftinterface}.

\section{Holography and complex CFTs}

As a model we consider the action of gravity coupled to  scalars with a potential. We specialize to the three dimensional bulk case for definiteness, since the  constructions on the CFT side we will present are for two dimensional CFTs. However, the holographic construction can   be straightforwardly generalized to any dimension.
\begin{align}\label{actiona}
    S= {1\over 16 \pi G_N}{\int} d^3x \sqrt{-g}\Big(R -{1\over 2} G_{ab}(\phi)\partial_\mu \phi^a\partial^\mu \phi^b- V(\phi)\Big)
\end{align}
The equations of motion following from this action are
\begin{align}\label{eqofma}
    R_{\mu\nu}-{1\over 2}g_{\mu\nu} R -{1\over 2} G_{ab}(\phi)\partial_\mu \phi^a \partial_\nu \phi^b +{1\over 4} g_{\mu\nu} G_{ab}(\phi)\partial_\rho\phi^a \partial^\rho \phi^b +{1\over 2} g_{\mu \nu }V(\phi)&=0\nonumber \\
    {1\over \sqrt{-g}}\partial_\mu \Big( \sqrt{-g} g^{\mu\nu}G_{ab}(\phi)\partial_\nu\phi^b\Big)- {1\over 2} {\partial G_{bc} \over \partial \phi^a} \partial_\mu \phi^b \partial^\mu \phi^c-{\partial V\over \partial \phi^a}&=0
\end{align}

The $AdS_3$ vacua in this model are critical points of the potential $V$ with negative vacuum energy.  The holographic renormalization group flow  \cite{deBoer:1999tgo,Freedman:1999gp} from  such a UV fixed point can be modeled  by a domain wall ansatz
\begin{align}\label{R2slicing}
   ds^2 = du^2 + e^{2A(u)}  (dx^2-dt^2), \quad \quad \phi^a=\phi^a(u)
\end{align}
corresponding to a spatially uniform flow.  With this ansatz the equations of motion (\ref{eqofma}) become a system of ordinary differential equations
\begin{align}\label{secondorder}
    {d^2A\over du^2}+2 \left({dA\over du}\right)^2+V(\phi)&=0\nonumber\\
{d^2\phi^a\over du^2} +2 {dA\over du} {d \phi^a\over du} + \Gamma^{a}_{\; bc} (\phi) {d\phi^b\over du} {d\phi^c\over du} -G^{ab} {\partial V\over \partial \phi^b}&=0   \nonumber \\
     \left({dA\over du}\right)^2-{1\over 4} G_{ab}{d \phi^a\over du} {d \phi^b\over du}  +{1\over 2} V(\phi)&=0
\end{align}
Here $\Gamma^{a}_{\; bc} (\phi)$ is the Christoffel connection derived from the metric $G_{ab}(\phi)$ appearing in the scalar kinetic term.

Since the scale factor $A$ can be interpreted as an energy scale,  the holographic beta function for an operator dual to the scalar field $\phi^a$ is defined as
\begin{align}\label{betaf1}
    \beta^a  = {d \phi^a \over d A}
\end{align}
A special class of flows can be derived from a superpotential $W(\phi)$  which is related to the potential by
\cite{Skenderis:1999mm}
\begin{align}\label{fake}
    V= 2\Big( G^{ab}(\phi) {\partial W \over \partial \phi^a} {\partial W\over \partial \phi^b}- W^2\Big)
\end{align}
The second order equations for the domain wall ansatz are then equivalent to the following first order system

\begin{align}\label{fake1}
     \frac{dA}{du} = W(\phi),\quad \frac{d\phi^a}{du} = -2G^{ab}\frac{\partial W}{\partial\phi^b}
\end{align}
Such equations arise in (gauged) supergravities where they are related to Killing spinor equations of domain walls that preserve some of the supersymmetries of the $AdS$ vacuum.  The validity  of (\ref{fake}) and (\ref{fake1}) is not dependent on supersymmetry and this is sometimes called "fake" supergravity \cite{Freedman:2003ax}. Using (\ref{fake1}) the beta function (\ref{betaf1}) can be expressed in terms of the superpotential as follows
\begin{align}\label{holobeta}
    \beta^a  = -2 G^{ab} {1\over W}{\partial W\over \partial \phi^b}
\end{align}
Note that the first order equations (\ref{fake1}) are less general than the full second order equations  (\ref{secondorder}), i.e. any solution of  (\ref{fake1})  is a solution of (\ref{secondorder})  but not vice versa. This can be interpreted as a fine tuning of the  source and expectation value in the flow. For our purposes of determining coefficients in the beta-function the first order expressions are sufficient.  Near a fixed point which we choose to be at $\phi^a=0$, the beta function takes in general the following form
\begin{align}\label{betageneral}
    \beta^a = -(2-\Delta^a) \phi^a +{\pi\over 2}\sum_{b,c} C^{a}_{\; bc}\phi^b \phi^c + \mathcal{O} (\phi^3)
\end{align}
where we assumed that the operators dual to $\phi^a$ have a definite scaling dimension (i.e. the mass matrix is diagonal).

\subsection{Holographic walking}
\label{sec:holowalk}

In this section we review a simple model adapted from \cite{Faedo:2019nxw} which illustrates the relation of complex CFTs and walking using the holographic RG-flow  described in the previous section. We  set $G_{ab}=\delta_{ab}$ and  furthermore limit ourselves to a single scalar $\phi$. Consider a superpotential with two critical points $\phi_1$ and $ \phi_2$ 
\begin{align}\label{dWdefa}
    {dW\over d\phi} = a (\phi-\phi_1) (\phi-\phi_2)
\end{align}
which gives 
\begin{align}\label{superpot}
    W= w_0 +  a \; \phi\Big( {1\over 3} \phi^2 - {1\over 2} \phi(\phi_1+ \phi_2) + \phi_1 \phi_2 \Big)
\end{align}
For real $\phi_{1,2}$ the critical points are IR and UV attractive and there can be an RG flow between them. When $\phi_1=\phi_2$ the critical points become degenerate and we are interested in the case where they move into the complex plane. For the superpotential to be real for real scalar field $\phi$ the critical points should be complex conjugate of each other and by a constant shift we choose the real part of the critical points to vanish
\begin{align}\label{criticala}
 \phi_1 = i \epsilon, \quad \phi_2 =\bar \phi_1= - i \epsilon   
\end{align}
When $\epsilon \ll 1 $ the complex critical points define conjugate complex CFTs which control the real "walking" RG evolution for a scalar  $\phi$ on the real axis close to the origin \cite{Gorbenko:2018dtm, Faedo:2019nxw,Faedo:2021ksi}.
Following \cite{Gorbenko:2018dtm} we consider the  critical points as defining two complex CFTs.  

Expanding the potential around the critical points $\phi=\phi_{1,2}+\delta \phi$
\begin{align}
    V  = V_0 + {1\over 2} m^2 \delta \phi^2 + \mathcal{O}(\delta\phi^3)
\end{align}
The central charge and the scaling dimension of the operator dual to the fluctuation of $\phi$ are given by\footnote{We work in units where $G_N=1$}
\begin{align}
    c = {3\over 2} \sqrt{-{2\over V_0}}, \quad \quad \Delta= 1+ \sqrt{1- {2m^2\over V_0} }
\end{align}
For $\epsilon\ll 1$, one gets for the two critical points
\begin{align}
    c_{1,2}= {3\over 2} {1\over |w_0|}\mp  i{ a \over w_0 |w_0|}\epsilon^3+\mathcal{O}(\epsilon^4), \quad \Delta_{1,2} = 2\mp i {4 a \over w_0}\epsilon -{8 a^2 \over 3 w_0^2}\epsilon^4+ \mathcal{O}(\epsilon^5)
\end{align}
Consequently, the holographic central charges of the two CFTs  get a small conjugate imaginary part. Scaling dimensions are very close to marginal, and the imaginary part leads to a spiraling flow described in \cite{Gorbenko:2018ncu,Faedo:2019nxw}. Note that the scaling dimension can also be determined by expanding  the holographic beta function (\ref{holobeta})
\begin{align}\label{betafunb}
    \beta_{1,2} = -\Big( \pm i{4 a\over w_0}\epsilon + {8 a^2\over 3 w_0^2} \epsilon^4+o(\epsilon^5)\Big) \delta \phi + \mathcal{O}(\delta\phi^2)
\end{align}
And the scaling dimension can be read off from the linear term by using (\ref{betageneral}).

\subsection{Holographic proof of the Im-flip relation}
In this section we show that the Im-flip relation for complex (conjugate) CFTs of \cite{Gorbenko:2018ncu} can be proved using the holographic beta function. We add a second scalar field $\psi$ which is  canonically normalized and choose the following superpotential
\begin{align}\label{phipsisupp}
    W= w_0 +  a\phi \Big( {1\over 3} \phi^2 - {1\over 2} \phi(\phi_1+ \phi_2) + \phi_1 \phi_2 \Big) + c_2 \psi^2 + c_3 \psi^2 \phi
\end{align}
For such a superpotential  (\ref{criticala}) together with $\psi=0$ will be a critical point. Other critical points are present but can be located far from the flow region of interest.  
We can determine the beta functions (\ref{holobeta})  for $\phi$ and $\psi$ near the critical points and expand to lowest nontrivial order in $\epsilon$ 
\begin{align}
    \beta_\phi &= \mp  i {4 a \epsilon\over w_0} \delta \phi -{2 a \over w_0} \delta\phi^2 -{2 c_3 \over w_0}\delta\psi^2 + \mathcal{O}(\epsilon^2) +\cdots \nonumber \\
    \beta_\psi&=\Big( -{4 c_2\over w_0} \mp i {4 c_3 \epsilon\over w_0} \Big) \delta \psi -{4 c_3 \over w_0} \delta \phi\delta\psi + \mathcal{O}(\epsilon^2)+\cdots
\end{align}
Where the signs correspond to the two conjugate critical points $\phi=\pm i \epsilon$. From (\ref{holobeta}) we can read off the OPE coefficients and imaginary parts  of the scaling dimension, giving
\begin{align}
    Im(\Delta_\phi)&= \mp {4 a\epsilon\over w_0} +\mathcal{O}(\epsilon^2), \quad Im(\Delta_\psi)= \mp {4 c_3 \epsilon\over w_0}+\mathcal{O}(\epsilon^2)\nonumber \\
   C_{\phi\phi\phi}&= -{1\over \pi } {4 a\over w_0}+\mathcal{O}(\epsilon^2), \quad \quad C_{\psi \psi\phi}= -{4\over \pi} {c_3\over w_0}+\mathcal{O}(\epsilon^2)
\end{align}
It is easy to see that the Im-flip relation is indeed satisfied up to second order in powers of $\epsilon$.
While the choice of superpotential  (\ref{phipsisupp}) looks non-generic, it captures all the important ingredients for coupling "walking" scalar where the complex conjugate critical points  are $\epsilon$ away from the real axis and a generic scalar $\psi$  coupled to $\phi$. The holographic Im-flip condition ceases to hold at order $\epsilon^3$, we note however that the original argument  in \cite{Gorbenko:2018ncu} relied on the complex fixed points being close to the real line and hence is also not expected to hold to that order. We also note that  the model of \cite{Faedo:2019nxw} had three critical points  of the superpotential at $\phi=0, \phi_1= \phi_r+ i\epsilon$ and $\phi_r - i\epsilon$. As long as the real location of the conjugate critical points $\phi_r$ is far from $\phi=0$, we can shift $\phi_r$ to zero and expand around the walking region, to leading order in $\epsilon$ and the fluctuations (where our above argument is valid) the beta functions will have the same form as above and the Im-flip relation follows.

\section{Complex conformal interfaces}

Conformal interfaces are constructed using a Janus-ansatz \cite{Bak:2003jk} which uses an $AdS_d$ slicing for a $d$-dimensional metric. For the $d=2$ case, which we focus on in this paper, one has 
\begin{align}\label{adsslice}
    ds^2 = du^2 + e^{2 A(u) } {dz^2- dt^2\over z^2}, \quad \phi=\phi(u)
\end{align}
With this ansatz the equations of motion (\ref{eqofma}) become ordinary differential equations for $A(u)$ and $\phi(u)$.
\begin{align}\label{eqofm}
    A''+2 (A')^2+e^{-2 A} + V(\phi)&=0\nonumber \\
    \phi''+2 A' \phi' - {\partial V\over \partial \phi}&=0
\end{align}
The $uu$ component of the gravitational equation is a constraint involving only first derivatives of $A$ and $\phi$.
\begin{align}\label{constr}
    (A')^2-{1\over 4} (\phi')^2+ e^{-2 A} + {1\over 2} V(\phi)&=0
\end{align}
In the Poincare sliced RG-flow ansatz (\ref{R2slicing}), where the holographic boundary is located at $u\to \infty$,  moving along the slicing coordinate is associated with RG-flow evolution from the UV to the IR. However, the $AdS_2$ slicing has a more complicated boundary structure, where the two boundary components $u\to \pm \infty $ correspond to two half spaces which are glued at a one dimensional interface which corresponds to the boundary of the $AdS_2$ space \cite{Bak:2003jk}. This  structure  can be seen by going to Fefferman-Graham coordinates \cite{Freedman:2003ax,Clark:2004sb}, but will not be needed in the present paper.

\subsection{Complex Janus}\label{janusSectionHolo}
The simplest Janus solution can be constructed using a single massless scalar which is dual to an exactly marginal operator in the CFT. In (\ref{actiona}) the potential is set to $V(\phi)=-2$ and the vacuum solution with constant scalar  is a unit radius $AdS_3$.   In this case the scalar equation of motion in (\ref{eqofm}) can be integrated once
\begin{align}
    \phi' = \alpha \; e^{-2 A}
\end{align}
and  solution of the equations of motion (\ref{eqofm}) and (\ref{constr}) can conveniently be expressed depending on the integration constant $\alpha$ \cite{Bak:2007jm}.
\begin{align}\label{janusfun}
    A(u) &= {1\over 2} \log\Big( {1\over 2} +{1\over 2}\sqrt{1-\alpha^2}\cosh(2u) \Big)\nonumber \\
    \phi(u)&= \phi_0 + \log \left({ 1+ \sqrt{1-\alpha^2} + \alpha\; \tanh u \over 1+ \sqrt{1-\alpha^2} -\alpha\;  \tanh u }\right)
\end{align}
The standard real Janus solution is obtained by choosing $\alpha$ real and the solution is non-singular if  $|\alpha|<1$. The holographic boundary corresponds to two half spaces at $u \to \pm \infty$ glued together at a one dimensional interface corresponding to the $AdS_2$ boundary \cite{Bak:2003jk,Clark:2004sb}

We will consider a complex interface solution by choosing $\phi_0=0$ and choosing $\alpha$ to be pure imaginary, i.e. $\alpha= i \;\gamma, \; \gamma \in \mathbb{R}$.  The asymptotic boundaries in the $AdS_2$ sliced geometry is reached when $u\to \pm \infty$ and the scalar approaches the values
\begin{align}
    \lim_{u\to \pm \infty} \phi(u) = {1\over 2} \log \left({1\pm i \gamma  \over 1\mp i \gamma}\right)
\end{align}
Since the asymptotic value of the massless scalar is related  to the source of a marginal $(h,\bar h)=(1,1)$ operator in the dual CFT this solution corresponds to turning on sources on the half space which are imaginary and conjugate of each other. Note that in this case the metric (\ref{janusfun}) remains  real. The parameter  $\gamma$ can take any real value. Note that the maximal imaginary distance of  the scalar in the complex Janus solution  is obtained as $\gamma\to \infty$  and is given by 
\begin{align}\label{imdistance}
  \Delta \phi= {1\over 2} (\phi|_{u=+\infty}-\phi|_{u=-\infty})={i\pi\over 2}
\end{align}
 which saturates  the $AdS_3$ wormhole bound of \cite{Maldacena:2026jqd,DiUbaldo:2026rly,Maldacena:Strings2026}. This is not surprising since AdS Janus interface solitons can be related to wormhole solutions by orbifolding the $AdS_2$ slice \cite{Ghodsi:2022umc}. In appendix \ref{app:c} we generalize this bound to the case where the massless scalars live on a non-linear sigma model with nontrivial metric, in this case the difference bound is replaced by a bound on the geodesic distance the scalars travel from $u=-\infty$ to $u=+\infty$ in a Janus flow.

The three dimensional holographic entanglement entropy is  determined by the Ryu-Takayanagi prescription \cite{Ryu:2006bv} 
\begin{align} \label{rtpres}
    S_{EE}({\cal A})  = {L[\Gamma_{\cal A}]\over 4 G_N}
\end{align} 
where $\Gamma_{\cal A}$ is a bulk geodesic anchored at the entanglement region ${\cal A}$ at the boundary.
There are two choices of  entangling region ${\cal A}$ for which the entanglement entropy in the presence of an interface can be readily calculated. For an entangling surface of length $2l$  which is symmetric about the interface at the origin \cite{Azeyanagi:2007qj,Chiodaroli:2010ur}. As shown in these papers, the minimal surface $\Gamma_{\cal A}$ is the  geodesic   at  constant $z=l$ and $u\in[-\infty,+\infty]$ and its length is given by
(\ref{rtpres}) by
\begin{align}
    L[\Gamma_{\cal A}]= u_{+\infty} -u_{-\infty} = 2\ln {2l\over \epsilon}- \ln {\sqrt{1+\gamma^2}}
\end{align}
where $\epsilon$ is a UV cutoff which is related to the Fefferman-Graham cutoff when the $AdS_2$ slicing is mapped to the Poincare slicing \cite{Chiodaroli:2010ur}.
Plugging  this into (\ref{rtpres}) the entanglement entropy becomes
\begin{align}
    S_{EE}({\cal A}) = {c\over 3} \log {2l\over \epsilon}-{c\over 12} \ln (1+\gamma^2) 
\end{align}
Where we use the relation for the central charge (at unit AdS radius)  on both sides of the interface 
\begin{align}\label{clr}
    c_L= c_R= c={3\over 2 G_N}
\end{align} The first term is the standard entanglement entropy for a CFT without a defect, the second term is called the defect g-factor or defect entropy.  We can expand the $g$ factor for small $\gamma$
\begin{align}
    \ln g= {c\over 3} \Big( -{1\over 4}\gamma^2 + {1\over 8} \gamma^4 +\mathcal{O}(\gamma^6)\Big)
\end{align}
Note that for the complex Janus both contributions are real due to the fact that the metric remains real under the complex Janus deformation.

A second entangling entropy is the one where the entangling surface  is at the interface. The holographic calculation for the   real Janus  was first performed in \cite{Gutperle:2015hcv} and the results can be used by replacing $\alpha \to i\gamma$
\begin{align}\label{EEatint}
    S_{EE}(A) = {c_{eff} \over 6} \ln {L\over \epsilon} = {1\over 4 G_N}e^{A(0)} \ln {L\over \epsilon} = {c\over 6} \sqrt{1+\sqrt{1+\gamma^2} \over 2}\ln{L\over \epsilon}
\end{align}
here $\epsilon$ is a FG UV-cutoff as before and $L$ is an IR cutoff for the semi infinite  entanglement  surface at the interface. Note that the central charge $c$ is replaced by an effective central charge which depends on the deformation. Again the result is real, however we note that the effective central charge is not bounded since $\gamma$ can take arbitrary large values.

A holographic formula for the reflection coefficient ${\cal R}$  and transmission   coefficient ${\cal T}$ was derived in \cite{Bachas:2022etu} 
\begin{align}
    {\cal T} ={2c_{LR} \over c_L+ c_R}, \quad {\cal R} = 1-{\cal T}
\end{align}
with 
\begin{align}
    c_{LR} = {3\over G_N}\Big(2 + 8\pi G_N \sigma\Big)^{-1}, \quad \quad  \sigma = \int_{-\infty}^\infty du \left( {d\phi\over du}\right)^2
\end{align}
For  a trivial interface we have $\sigma=0$ and ${\cal T}=1$ and ${\cal R}=0$, i.e. complete transmission without reflection. For the complex Janus solution we find
\begin{align}\label{sigmares}
    \sigma= {8\over \gamma}\tan^{-1}\left(\gamma\over 1+\sqrt{1+\gamma^2}\right) -4
\end{align}
It is easy to verify that for any nonzero gamma ${\cal T} > 1$ and ${\cal R}<0$ which is impossible for a unitary CFT and a consequence of dealing with a complex CFT which is not unitary. It is instructive to expand ${\cal T}$ for small values of $\gamma$
\begin{align}
    {\cal T}= 1+ {8 \pi \gamma^2\over c}- {8\pi (3c -40 \pi)\gamma^4\over 5 c^2 }+ \mathcal{O} (\gamma^6)
\end{align}

 \subsection{Complex RG-flow interfaces}
 For the complex Janus interface of the previous section the massless scalar takes  imaginary values.  Since it appears quadratically in the Einstein's equations the metric and quantities derived from it such as the central charge remain real.  In order to construct holographic interface solutions where the CFTs on both sides have complex conjugate central charges we have to consider scalars with potentials and so called RG-flow interfaces \cite{Gaiotto:2012np}  and their holographic realizations (see e.g. \cite{Arav:2020asu,Chen:2021mtn}).
 
For simplicity, we consider a single scalar field  with a potential $V(\phi)$ which has two extrema which are complex conjugates.
 \begin{align}
    \left. {\partial V\over \partial \phi}\right|_{\phi=\phi_\pm} =0, \quad \quad (\phi_\pm)^* = \phi_{\mp}
 \end{align}
 If the scalar takes the conjugate extremal value there are two  complex conjugate AdS solution of the equations of motion (\ref{eqofm}) and (\ref{constr}) given by
 \begin{align}
    A(u) &= \ln\Big[\cosh\Big( {u\over L_{\pm}}\Big)\Big] +\ln L_{\pm}, \quad \quad \phi(u) = \phi_{\pm}
\end{align}
Where the $AdS$ radius is fixed by the value of the potential at the extrema $V_{\pm}= V(\phi_{\pm})$
\begin{align}
    L_{\pm} &= \sqrt{-{2\over V_\pm}}, \quad \quad (L_{\pm})^*=L_{\mp}
\end{align}
  The holographic central charge of the two conjugate vacua are complex conjugate due to the relation $\bar L_{\pm}= L_{\mp}$
\begin{align}
    c_{\pm}= {3L_{\pm}\over 2 G_N}
\end{align}
We would like to construct an RG-flow interface where the scalar field approaches the conjugate extrema $\phi\to \phi_\pm $ as $u\to \pm \infty$. Unlike the Janus solution, we have to integrate the second order equations of motion (\ref{eqofm}) and (\ref{constr}) numerically. In order to determine the initial conditions at some finite but large positive $u$ we linearize around one extrema, $\phi= \phi_+ +\delta \phi$. The linearized equation becomes 
\begin{align}
    \delta \phi'' + 2 a' \delta \phi'-m_+^2 \delta \phi =0
\end{align}
where the mass is determined by expanding the potential around $\phi_+$. Since the scalar appears quadratically in the Einstein equations to linear order in the fluctuations, the backreaction on the metric can be neglected, and the metric factor for large positive  $u$ behaves as
\begin{align}
       A(u) \sim    {u\over L_+}+ \ln \left({  L_+\over 2 }\right) +  A_0+ A_1 e^{- {2u\over L_+}}\cdots 
\end{align}
and the linearized scalar is given by
\begin{align}\label{linfluc}
 \delta\phi= c_1 e^{ -\Delta { u\over L_+}   }+ c_2 e^{ -(2-\Delta) { u\over L_+}   }+\cdots   
\end{align}
where the scaling dimension is given by the standard holographic formula
\begin{align}\label{scaledel}
    \Delta =1+\sqrt{ 1+{  m^2 L_+^2}}
\end{align}
and sub-leading terms  fall off faster with larger exponential factors for large $u$. Note that for complex extrema the scaling dimension  (\ref{scaledel}) will in general be complex. If the real part satisfies $0<\Re(\Delta)<2$, we call it "UV attractive" as both terms in the linearized scalar fluctuation (\ref{linfluc}) decrease as $u$ increases, otherwise we call the extrema "UV-repulsive". In the following we are interested in RG-flow interfaces between two UV attractive extrema.

The initial conditions are determined at a $u=u_0$ which is large enough such that the linearization is a good approximation. The constraint equation (\ref{constr}) can be used to eliminate one of the four parameters $c_1,c_2, A_0, A_1$ and the linearized form can be used to parameterize the initial conditions for $\phi, \phi'$ and $A, A'$ at $u=u_0$ and integrate the second order equations of motion (\ref{eqofm}).
Evaluating the constraint (\ref{constr}) along the solution is a check for the accuracy of the numerical solution. In general we have to  vary the initial condition to shoot for the solutions corresponding to the RG-flow interface.
In the next section   we use  model of a single scalar  related to the discussion in section \ref{sec:holowalk}.
\subsubsection{Quartic superpotential}
The first example is a toy model of a single scalar with a quartic  superpotential, which was first discussed in \cite{Faedo:2019nxw} defined by\footnote{The reason why we do not use the   superpotential (\ref{dWdefa}) used in the im-flip argument is that the additional real UV fixed point allows a better control over the numerics.}
 \begin{align}\label{cubic}
    {dW\over d\phi} = a \; \phi  (\phi-\phi_r- i\epsilon) (\phi-\phi_r+ i\epsilon)
\end{align}
The superpotential has a real critical point at $\phi=0$ and two complex conjugate critical points at $\phi_\pm =\phi_r\pm i \epsilon$  as discussed in section \ref{sec:holowalk} for this potential to describe walking we take the imaginary part $\epsilon$ to be small. The superpotential is then given by
\begin{align}\label{wdef}
   W=  w_0 + a\left(\frac{\phi^4}{4} - \frac{2\phi_r\phi^3}{3} + \frac{(\phi_r^2+\epsilon^2)\phi^2}{2}
   \right)
\end{align}
The potential is determined by (\ref{fake}), and the value of the potential at the two critical points is given by
\begin{align}
    V_\pm=V(\phi_\pm) &= - {1\over 72}\Big(-12 w_0 - a (\phi_r\pm i \epsilon)^3 (\phi_r\mp i  3 \epsilon)\Big)^2 \nonumber \\
    &=-2\left(w_0 + \frac{a\;\phi_r^4}{12}\right)^2 - 2\left[a\;\phi_r^2\left(w_0 + \frac{a\phi_r^4}{12}\right)\right]\epsilon^2 \mp\frac{2i}{9}a\;\phi_r^4(12w_0 + a\phi_r^4)\epsilon^3 + \mathcal{O}(\epsilon^4)
\end{align}
Hence the AdS vacuum corresponding to the scalar at the extrema $\phi=\phi_\pm$ has a complex cosmological constant and hence an AdS metric determined with complex conjugate $AdS$ radii
\begin{align}
    L_{\pm} &= \sqrt{-{2\over V_\pm}}\nonumber \\
    &= {12 \over |a \phi_r^4+12 w_0|}\left( 1 - \epsilon^2\frac{6   a \phi_r^2} {a \phi_r^4+12 w_0}
    \mp \epsilon^3 \frac{8 i a \phi_r }{a \phi_r^4+12 w_0}+ \mathcal{O}(\epsilon^4)\right)
\end{align}
The scaling dimensions for the fluctuation around the conjugate extrema which are needed for setting up the initial conditions as described in the previous section are
\begin{align}
  \Delta_{\pm} &=2 \mp i  {48 a \phi_r \over a \phi_r^4 +12 w_0} \epsilon +  {48 a\over a \phi_r^4 +12 w_0  }\epsilon^2+\mathcal{O}(\epsilon^3)
\end{align}

Due to the imaginary terms in the scaling dimensions $\Delta_{\pm}$, the flows display  a spiraling behavior once near one of the complex fixed points. In addition if  the real parts satisfy  $\Re(\Delta_{-})>0, \Re(\Delta_-)<2$ flow spirals in to the complex fixed points as $u\to \pm \infty$, i.e. is UV attractive.

Note that it follows from (\ref{fake}) that the extrema of the superpotential $W$ are extrema of the potential $V$. There are in general additional extrema of $V$ which can be  real extrema with positive or negative $V_0$, or complex conjugate ones.  Since our goal is to illustrate RG-flow interfaces between $\phi_{\pm}$ we choose the  parameters in (\ref{wdef}) such that the real extrema with positive $V$ are located  far away from $\phi_{\pm}$. In addition we demand that any additional real extrema is UV repulsive and satisfies the Breitenlohner Freedman bound as well as any additional complex conjugate extrema to be UV repulsive. This means that the only RG-flow interfaces which can occur are  $\phi_\pm \leftrightarrow \phi_\mp$ and $\phi_\pm \leftrightarrow \phi_{\pm}$. A choice of parameters which satisfies these criteria is given by
\begin{align}
    \phi_\pm &= 1\pm  0.1\; i, \quad a=-0.8, \quad w_0=1
 \end{align}

\begin{table}[htbp]
\centering
\renewcommand{\arraystretch}{1.25} 
\begin{tabular}{cccc}
\hline
\hline
&$\phi_*$ & $V(\phi_*)$&$\Delta$ \\
\hline
1.&$1.000000 + 0.100000 i$ & $-1.72739+0.001983 i$ &$1.965370 + 0.344306 i$ \\
2.&$1.000000 - 0.100000 i$ & $-1.72739-0.001983 i$&$1.965370 - 0.344306 i$ \\
3.&$0$ & -2 &$3.616000$ \\
4.&$0.678925 + 0.513162 i$ & $-1.69356+0.0558607 i$&$3.021537 + 0.192991 i$ \\
5.&$0.678925 - 0.513162 i$ & $-1.69356-0.0558607 i$&$3.021537 - 0.192991 i$ \\
6.&$-2.938599$ &749.163 & $3.911450$ \\
7.&$4.247415$ &710.812 & $3.922455$ \\
\hline
\hline
\end{tabular}
\caption{Fixed points, cosmological constant and  scaling dimensions for $a = -0.8, W(0) = 1$, $\phi_+ = 1 + 0.1i$}
\label{tab:scaling_values_numerics}
\end{table}
All extrema of the potential $V$ and the associated cosmological constant and scaling dimensions of the scalar fluctuation are listed in Table \ref{tab:scaling_values_numerics}.

\begin{figure}[H]
    \centering

    \begin{subfigure}[b]{0.32\textwidth}
        \centering
        \includegraphics[width=\linewidth]{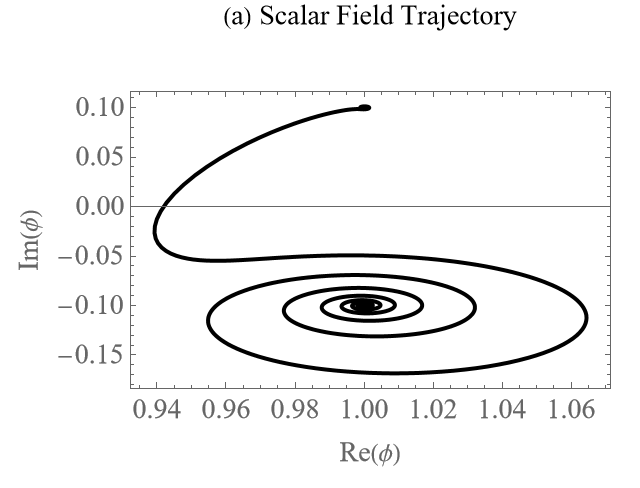}
        \label{fig:flow1}
    \end{subfigure}\hfill
    \begin{subfigure}[b]{0.32\textwidth}
        \centering
        \includegraphics[width=\linewidth]{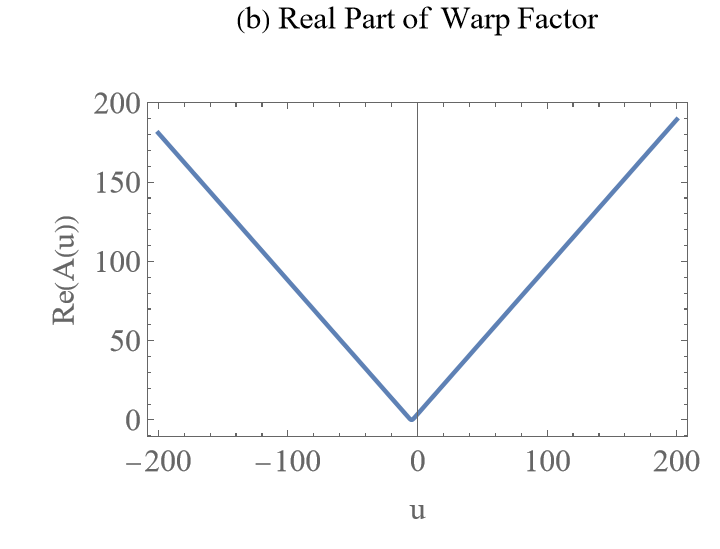}
        \label{fig:flow1a}
    \end{subfigure}\hfill
    \begin{subfigure}[b]{0.32\textwidth}
        \centering
        \includegraphics[width=\linewidth]{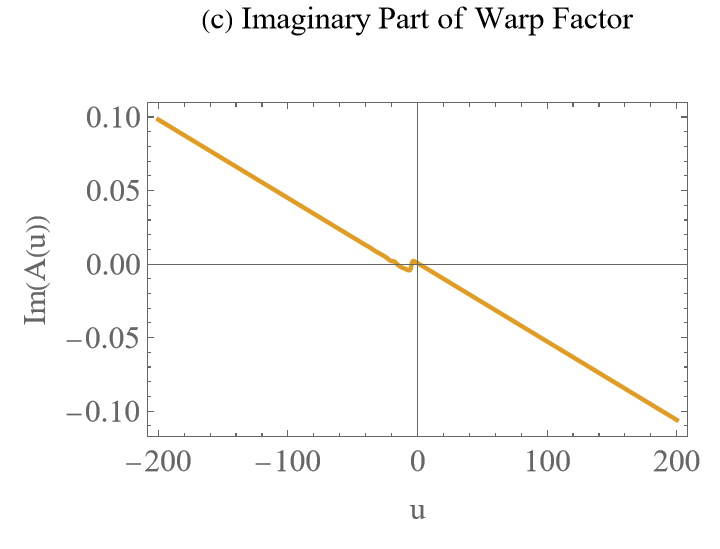}
        \label{fig:flow1b}
    \end{subfigure}

    \centerline{
    \begin{tabular}{lll}
     $\phi(0) = 0.999527 + 0.0983055 i$ & \hspace{1cm} & $A(0) = 3.63435 + 0.000712791 i$ \\
     $\phi'(0) = 0.000637307 - 0.000100837 i$ & & $A'_{\text{calc}}(0) = 0.928978 - 0.000532793 i$
    \end{tabular}}


    \begin{subfigure}[b]{0.32\textwidth}
        \centering
        \includegraphics[width=\linewidth]{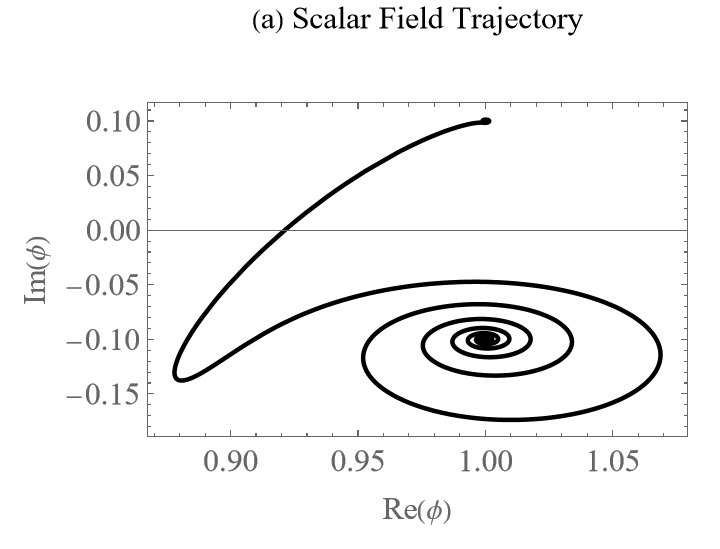}
        \label{fig:flow4}
    \end{subfigure}\hfill
    \begin{subfigure}[b]{0.32\textwidth}
        \centering
        \includegraphics[width=\linewidth]{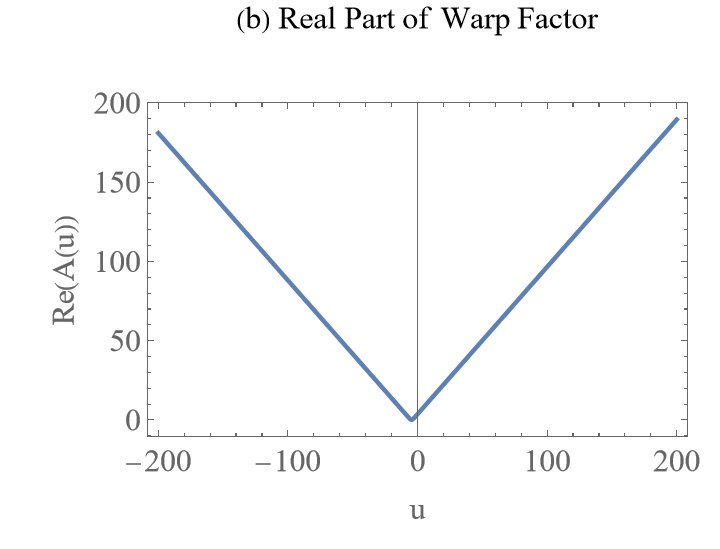}
        \label{fig:flow4a}
    \end{subfigure}\hfill
    \begin{subfigure}[b]{0.32\textwidth}
        \centering
        \includegraphics[width=\linewidth]{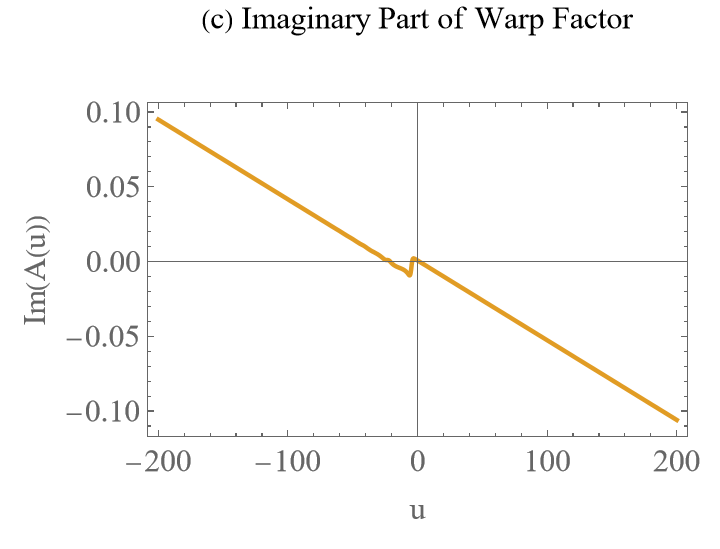}
        \label{fig:flow4b}
    \end{subfigure}

    \centerline{
    \begin{tabular}{lll}
     $\phi(0) = 0.999527 + 0.0983055 i$ & \hspace{1cm} & $A(0) = 3.63435 + 0.000712791 i$ \\
     $\phi'(0) = 0.000708307 - 0.000100837 i$ & & $A'_{\text{calc}}(0) = 0.928978 - 0.000532795 i$
    \end{tabular}}

    \caption{Flows for the superpotential defined by (\ref{cubic}). Top row: baseline initial conditions. Bottom row: identical initial conditions except for the perturbation to $\phi'(0)$ shown above each row. Both examples show flows between two complex conjugate points}
    \label{fig:group1}
\end{figure}

Note that the extrema 3,4 and 5 are UV-repulsive and the dS vacua 6 and 7 are far away from the relevant extrema 1 and 2 and an artifact of the choice of a model potential.

\begin{figure}[H]
    \centering

    \begin{subfigure}[b]{0.32\textwidth}
        \centering
        \includegraphics[width=\linewidth]{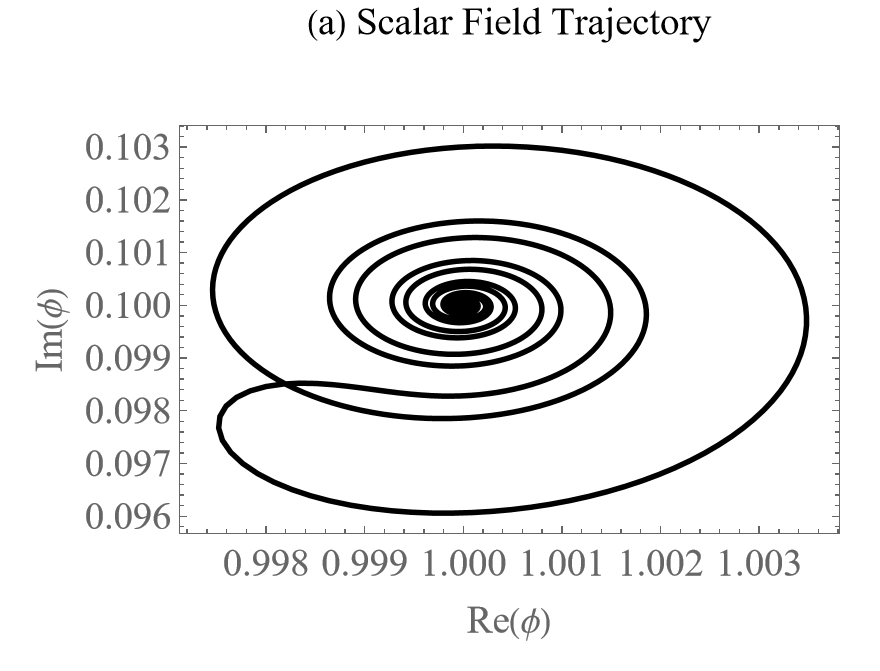}
        \label{fig:flow2}
    \end{subfigure}\hfill
    \begin{subfigure}[b]{0.32\textwidth}
        \centering
        \includegraphics[width=\linewidth]{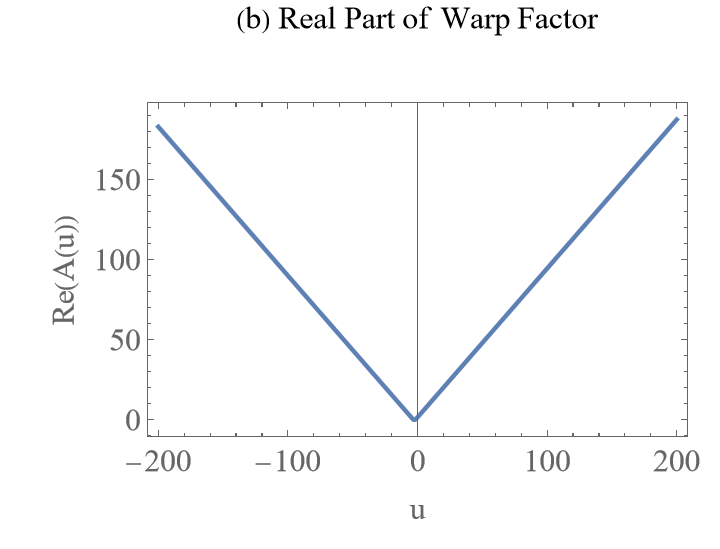}
        \label{fig:flow2a}
    \end{subfigure}\hfill
    \begin{subfigure}[b]{0.32\textwidth}
        \centering
        \includegraphics[width=\linewidth]{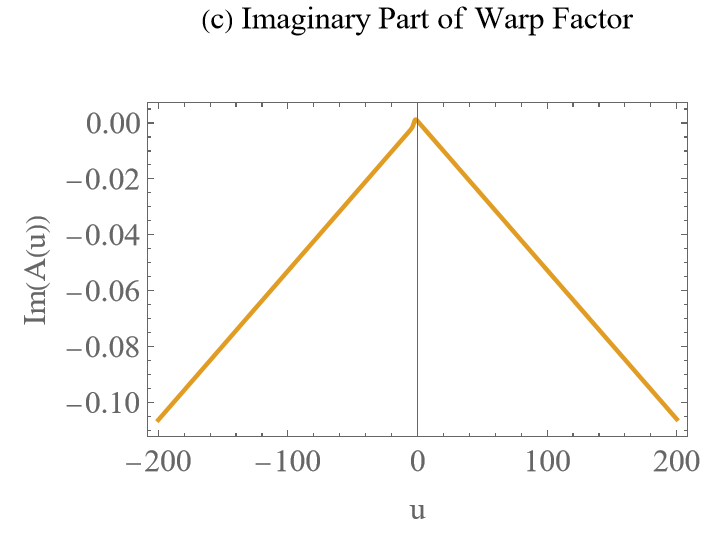}
        \label{fig:flow2b}
    \end{subfigure}

    \vspace{4pt}
    \centerline{
    \begin{tabular}{lll}
     $\phi(0) = 0.999527 + 0.0983055 i$ & \hspace{1cm} & $A(0) = 1.63435 + 0.000712791 i$ \\
     $\phi'(0) = 0.000637307 - 0.000100837 i$ & & $A'_{\text{calc}}(0) = 0.908648 - 0.000515408 i$
    \end{tabular}}

    \vspace{12pt}

    \begin{subfigure}[b]{0.32\textwidth}
        \centering
        \includegraphics[width=\linewidth]{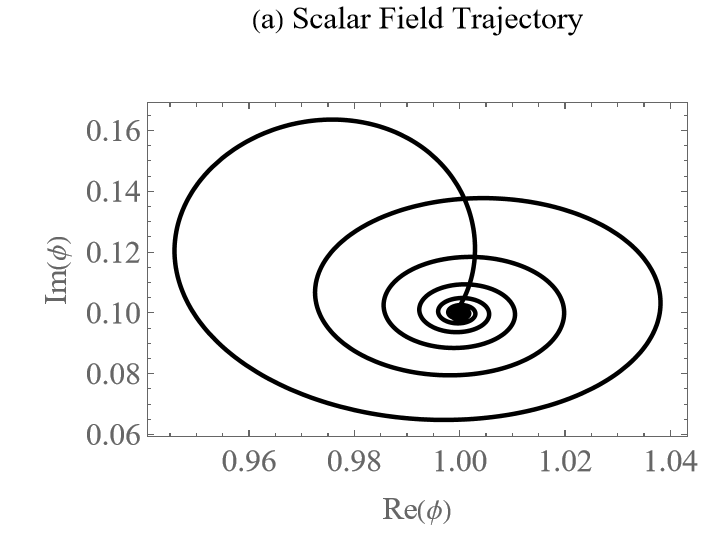}
        \label{fig:flow3}
    \end{subfigure}\hfill
    \begin{subfigure}[b]{0.32\textwidth}
        \centering
        \includegraphics[width=\linewidth]{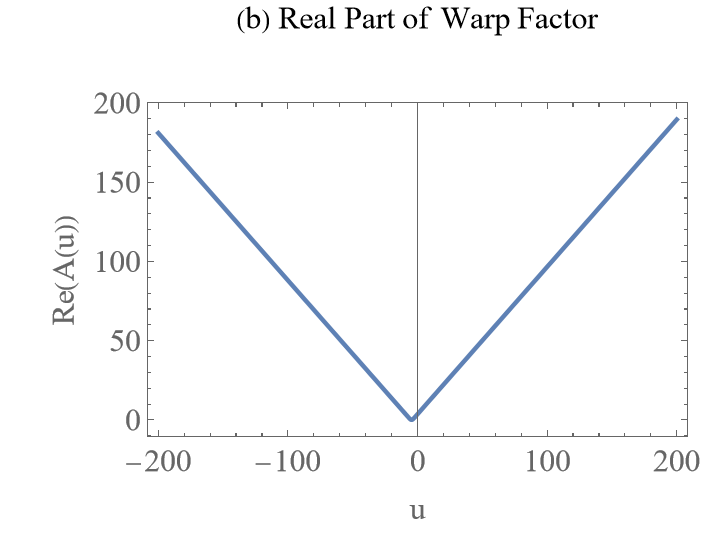}
        \label{fig:flow3a}
    \end{subfigure}\hfill
    \begin{subfigure}[b]{0.32\textwidth}
        \centering
        \includegraphics[width=\linewidth]{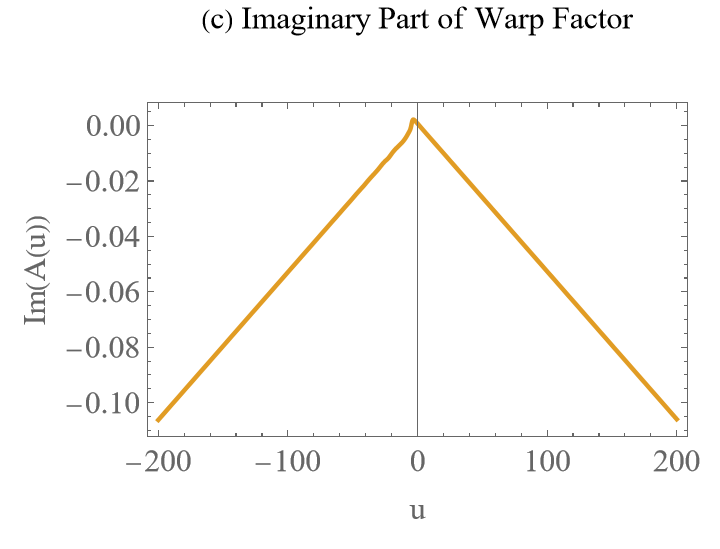}
        \label{fig:flow3b}
    \end{subfigure}

    \vspace{4pt}
    \centerline{
    \begin{tabular}{lll}
     $\phi(0) = 0.999527 + 0.0979215 i$ & \hspace{1cm} & $A(0) = 3.63435 + 0.000712791 i$ \\
     $\phi'(0) = 0.000637307 - 0.000100837 i$ & & $A'_{\text{calc}}(0) = 0.928978 - 0.000532675 i$
    \end{tabular}}

    \caption{Flows for the superpotential defined by (\ref{cubic}). Top row: $A(0)$ real part lowered to $1.63435$. Bottom row: baseline $A(0)$, with $\mathrm{Im}\,\phi(0)$ perturbed as shown above the row. Both examples are for flows from and to the same critical complex point.}
    \label{fig:group2}
\end{figure}

The plots in figure \ref{fig:group1}  represent a  RG-flow interfaces between the complex conjugate vacua. The plots in figure \ref{fig:group2} represent a RG-flow interface with vacuum 1 on both sides.  The flows are also very sensitive to the initial conditions as is illustrated by the two examples of each kind which only differ by a small change in the initial conditions.

\section{Complex CFT interfaces}
\label{complexcftinterface}
In this section we will consider two dimensional conformal interfaces with complex CFTs which are conjugate to each other on both sides of the interface.

\subsection{Complex Janus CFT } 

In the previous sections, we noted that the Janus solution is dual to interface CFTs with equal real central charges on either side.  Therefore, the non-unitarity in the complex Janus CFT must come from the imaginary values of coupling constants instead of any complex values in the spectrum of the theory. 

A natural CFT realization is a noncompact free boson $X$ with a
different  kinetic coupling or "stiffness" on either side of the interface \cite{Baig:2024}.
\begin{align}\label{janus_action}
    S = k_L \int_{x<0} \eta^{\mu\nu}\partial_{\mu}X\partial_{\nu}X+k_R \int_{x>0} \eta^{\mu\nu}\partial_{\mu}\tilde{X}\partial_{\nu}\tilde{X}
\end{align}
The holographic dictionary identifies the asymptotic scalar values $\phi_{\pm}$
with the boundary marginal couplings, so the complex Janus deformation sets
\begin{align}\label{janus_couplings}
    k_{L} = k_0\,e^{-i\arctan\gamma_{cft}}\,, \qquad
    k_{R} = k_0\,e^{+i\arctan\gamma_{cft}} = \overline{k_{L}}\,,
\end{align}
for $k_0\in \mathbb{R}$.  The central charge of a free boson is
$c = 1$ independent of the kinetic normalization, since changing  $k$ corresponds to an  exactly marginal deformation.
Hence $c_L = c_R = 1$, which is consistent with the holographic dual
(\ref{clr}).  The two half-space theories are the same free boson analytically continued to
imaginary coupling.\footnote{See \cite{Kawamoto:2025oko,Harper:2025lav} for previous discussion  of the imaginary Janus CFT interface and its relation to traversable wormholes in AdS$_3$ and \cite{Maeda:2026awj} for a discussion of an imaginary deformed BCFT and its holographic dual. }

\subsubsection{Reflection and transmission}

Varying (\ref{janus_action}) yields the wave equation $k\,\Box X = 0$ in each
half-space and, from the boundary term at $x=0$, the matching conditions
\begin{align}\label{janus_bcs}
  X\big|_{0^{-}}=X\big|_{0^{+}} ,
    \qquad
    k_L\,\partial_x X\big|_{0^{-}} = k_R\,\partial_x X\big|_{0^{+}}\,.
\end{align}
The first condition imposes  continuity of the field, while the second follows from the vanishing of the total derivative terms for the variation of the action at $x=0$.

Here we follow a similar argument that is used in \cite{Bachas_2002} to calculate the reflection and transmission coefficients while allowing for complex coefficients. For a right-moving mode $e^{-i\omega(t-x)}$ incident from $x<0$, the
ansatz
\begin{align}
    X = \bigl(e^{-i\omega(t-x)} + {R}\,e^{-i\omega(t+x)}\bigr)\Theta(-x)
      + {T}\,e^{-i\omega(t-x)}\,\Theta(x)
\end{align}
and imposing the conditions (\ref{janus_bcs}) determines the coefficients $R$ and $T$ 
\begin{align}\label{tr_amplitudes}
    T = \frac{2k_L}{k_L + k_R}\,, \qquad R = \frac{k_L - k_R}{k_L + k_R}\,.
\end{align}

Comparing the ratio of energy flow ${k\over 2} \partial_t X \partial_x X$ for the reflected and transmitted wave to the incident wave we can determine the energy reflection and transmission coefficients

\begin{align}
   \mathcal{R}&=R^2
      = \left(\!\frac{k_L - k_R}{k_L + k_R}\!\right)^{\!2}, \qquad  \mathcal{T}= {k_R\over k_L} T^2= \frac{4k_L k_R}{(k_L + k_R)^{2}}
\end{align}
Which satisfy $\mathcal{R}+\mathcal{T}=1$ as implied by energy conservation.

The transmission and reflection coefficients for the stress tensor are thus given by
\begin{align}\label{TR_janus_cft}
    \mathcal{T}
      = \frac{4k_L k_R}{(k_L + k_R)^{2}}\,,
    \qquad
    \mathcal{R}
      = \left(\!\frac{k_L - k_R}{k_L + k_R}\!\right)^{\!2}
      = 1 - \mathcal{T}\,.
\end{align}
The identity $\mathcal{T} + \mathcal{R} = 1$ holds for all values of $k_L$ and
$k_R$, real or complex, because it follows from the algebraic identity
$(k_L - k_R)^2 + 4k_L k_R = (k_L + k_R)^2$.  It is a consequence of the
conformal Ward identity of the folded theory and is insensitive to whether the
gluing is unitary or not.

Substituting the complex Janus couplings (\ref{janus_couplings}), one finds
$k_L k_R = k_0^2$ which is real and $k_L + k_R = 2k_0\cos(\arctan\gamma_{cft})$, also real. So,
\begin{align}\label{TR_result_cjanus}
    \mathcal{T}
      = \sec^{2}(\arctan\gamma_{cft}) = 1 + \gamma_{cft}^{2} \;\geq 1\,,
    \qquad
    \mathcal{R}
      = -\tan^{2}(\arctan\gamma_{cft}) = -\gamma_{cft}^{2} \;\leq 0\,.
\end{align}
Both coefficients are real for all $\gamma_{cft}$, and both become one and zero,
respectively, at $\gamma_{cft} = 0$.  The super-transmission $\mathcal{T} > 1$ and
negative reflection $\mathcal{R} < 0$ arise because the imaginary phase of the
couplings reduces the real part of $k_L + k_R$ from $2k_0$ to
$2k_0/\sqrt{1+\gamma_{cft}^2}$, while leaving the product $k_L k_R = k_0^2$
unchanged, thereby increasing  the ratio in (\ref{TR_janus_cft}).  Physically,
the imaginary marginal coupling localised at $x = 0$ acts as a non-Hermitian
source that redirects energy from the reflected channel into the transmitted
channel.  This is consistent with the holographic observation that the metric
remains real throughout the complex Janus geometry: the stress tensor on both
sides retains a real structure.

The result (\ref{TR_result_cjanus}) is in qualitative agreement with the
holographic computation of the section \ref{janusSectionHolo}. Expanding for small
$\gamma_{cft}$,
\begin{align}\label{janusT}
    \mathcal{T} = 1 + \gamma_{cft}^{2} + \mathcal{O}(\gamma_{cft}^{4})\,,
\end{align}
compared to the holographic result
$\mathcal{T}^{\rm holo} = 1 + 8\pi\gamma_{cft}^{2}/c + O(\gamma_{cft}^{4})$.  Both predict
super-transmission at order $\gamma_{cft}^{2}$. Note that $\gamma_{cft}$ and $\gamma$ as used in the holographic model in the previous section are related due to the duality $\phi \leftrightarrow g$. This gives $\gamma_{cft} = N_{\gamma}\gamma$ for some normalization $N_{\gamma}$, which results in 
\begin{equation}
    \gamma_{cft} = \sqrt{\frac{8\pi}{c}}\gamma
\end{equation}
Thus we see agreement in the transmission coefficient up to quadratic order in $\gamma$.

\subsubsection{Folded boundary state and g-factor}

The interface admits a boundary state description via the folding trick \cite{Oshikawa:1996dj,Bachas_2002}.  Denoting $X_{L,R}$ as the scalar field for $x<0$ and $x>0$ respectively and rescaling $X_{L,R} \to  (k_{L,R})^{-{1\over 2}} X_{L,R}$ to obtain a canonically normalized stress tensor, the boundary conditions (\ref{janus_bcs}) can be expressed in terms of the currents $\partial_{\pm} X_{L,R}=(\partial_t\pm \partial_x)X_{L,R}$
\begin{align}\label{janus-bc2}
    \partial_{+} X_{L}\mid_{x=0^-}&= {1\over 2} {k_L+k_R\over \sqrt{k_L k_R}} \partial_+X_R \mid_{x=0^+} + {1\over 2} {k_L-k_R\over \sqrt{k_L k_R}} \partial_-X_R\mid_{x=0^+}\nonumber \\
     \partial_{-} X_{L}\mid_{x=0^-}&= {1\over 2} {k_L-k_R\over \sqrt{k_L k_R}} \partial_+X_R \mid_{x=0^+} + {1\over 2} {k_L+k_R\over \sqrt{k_L k_R}} \partial_-X_R\mid_{x=0^+}
\end{align}
The folding trick 
relates an interface between $CFT_1$ on a half  space $x<0$  and $CFT_2$  on the half space  $x>0$ to a boundary state in the tensor product $\overline{CFT_1}\otimes CFT_2$. We identify $X_1(x,t) = X_L(-x,t)$ and $X_2(x,t)= X_R(x,t)$ and  the conditions (\ref{janus-bc2}) are imposed by a boundary state acting on currents
\begin{align}
    \left.  \Big(\partial_+ X^{(i)} - S^{ij} \partial_- X^{(j)}\Big)\right|_{x=0}\mid B\rangle \rangle=0
\end{align}
with the gluing matrix given by 
\begin{align}\label{S_janus}
    S = \frac{1}{k_L + k_R}
    \begin{pmatrix}
        k_L - k_R & 2\sqrt{k_L k_R} \\[4pt]
        2\sqrt{k_L k_R} & k_R - k_L
    \end{pmatrix}= \begin{pmatrix}
        -\cos(2\theta) & \sin(2\theta)\\
        \sin(2\theta ) & \cos(2\theta)
    \end{pmatrix}
\end{align}

Using the mode expansion of the bosons and conventions given in appendix \ref{app:a1} the boundary state is given by
\begin{align}
    |B\rangle\rangle
      =  {1\over \sqrt{\sin 2\theta}} \exp\!\Biggl(
          \sum_{n=1}^{\infty}
          \frac{1}{n}S_{ij}\,\alpha^{(i)}_{-n}\bar{\alpha}^{(j)}_{-n}
        \Biggr) 
        \int dk e^{i k( y^1 \sin{\theta}+ y^2\cos \theta )} \mid  k\; \sin\theta \rangle_{(1)} \otimes \mid k\; \cos\theta \rangle_{(2)} 
\end{align}
The integration is over momenta in the Dirichlet direction, and the normalization is obtained from taking a limit for the compact boson case described in appendix \ref{app:a1}.  It follows that the interface entropy, or g-factor (which is determined from the constant term in the entanglement entropy which is symmetric about the interface), is given by
\begin{align}\label{janusgfactor}
     g = \langle 0\mid B\rangle \rangle = { {1\over \sqrt{\sin 2\theta}} = \sqrt{  k_L+k_R \over 2\sqrt{k_L k_R}}}
\end{align}
We also quote the result for the CFT calculation of the entanglement entropy at the interface, which can be obtained by analytic continuation of the result for a Janus CFT \cite{Sakai:2008tt, Brehm:2015lja,Gutperle:2015hcv}.
The CFT result is of the same form as (\ref{EEatint}) with
\begin{align}
c_{eff} = c\sqrt{1+\sqrt{1+\gamma_{cft}^2} \over 2}\
\end{align}

\subsection{Complex linear dilaton interface }
As a second example  we consider  linear dilaton conformal field theories, which is a noncompact boson with a background charge $Q$. On a general Riemann surface the action is given by
\begin{align}\label{lineardil}
    S  &= \frac{k}{4\pi}\int d^2 \sigma \sqrt{g } \, \partial_\alpha  X \partial^\alpha  X+ {1\over 4\pi}  \int d^2 \sigma \sqrt{g} Q X R^{(2)}[g]
\end{align}
We choose a canonically normalized boson (or alternatively rescale the noncompact boson to set $k=1$) and the theory is determined by the background charge.  The second term in (\ref{lineardil}) vanishes on the plane or cylinder, however the stress energy tensor is modified by the presence of a background charge. We review our conventions and some facts about the linear dilaton CFT in appendix  \ref{app:a2}.

$\text{CFT}_1$ and $\text{CFT}_2$,  are chosen to be complex conjugates and the fields $\phi_1$ and $\phi_2$ have noncompact target spaces. In a complex linear dilaton theory, the background charges possess both a real and  imaginary parts\footnote{A purely imaginary background charge can be related to a boson with a timelike kinetic term by a rescaling and is sometimes called timelike linear dilaton theory.} and are complex conjugate
\begin{align}
    Q^{(1)} &= q_R + i q_I, \quad\quad   Q^{(2)} = q_R - i q_I
\end{align}
The central charge of an individual complex linear dilaton theory is explicitly complex, reflecting its underlying non-hermiticity:
\begin{align}
    c_{1,2} = 1 +6(Q^{(1,2)})^2 = 1+ 6(q_R^2 - q_I^2) \pm  12i q_R q_I
\end{align}
However, because the two theories are complex conjugates, the total central charge of the folded product theory $\text{CFT}_1 \otimes \overline{\text{CFT}_2}$ is real:
\begin{align}
    c_{\text{total}} = c_1 + c_2 = 2 +12(q_R^2 - q_I^2)
\end{align}
This is a similar setup that was used for the complex Liouville string worldsheet theory \cite{Collier:2024kmo,Collier:2024kwt,Collier:2024mlg}. In the present paper we are not restricting the critical charge and we consider the noncompact linear dilaton theory instead of an action with an exponential  Liouville term added to  (\ref{lineardil}).

Because the bosons are  noncompact, the continuous momentum eigenvalue is given by $k \in \mathbb{R}$. The primary vertex operators $V^{(i)}_k = :e^{ik\phi^{(i)}}:$, for $ i=1,2$ have anomalous scaling dimensions shifted by the complex background charge:
\begin{align}\label{scale_complex_dilaton}
    \Delta^{(1,2)}_k = h + \bar{h} = 2 \left( \frac{k^2}{4} + \frac{iQ^{(1,2)}}{2} k \right) = \left(\frac{k^2}{2} \mp q_I k\right) + i q_R k
\end{align}
The appearance of the linear shift $\mp q_I k$ in the real part of the scaling dimension shows that the theory circumvents standard unitary lower bounds, a known feature of complex Liouville and non-Hermitian systems. As is standard due to the linear dilaton inner product, we can make the shift $k\to k-iQ^{(1,2)}$, which then gives us the known expression 
\begin{equation}
    \Delta_k^{(1,2)} = \frac{1}{2}\left(k^2 + (Q^{(1,2)})^2\right)\\
\end{equation}
This is the expected result for a unitary theory with real background charge, and further has the desired complex conjugate property for complex $Q^{(i)}$.

The modes of the stress-energy tensor are then given by 
\begin{align}
    L_m^{(i)} = \frac{1}{2}\sum_n a_{m-n}^{(i)}a_n^{(i)} + \frac{ iQ^{(i)}}{\sqrt{2}}(m +1)\; a_m^{(i)} 
\end{align}
With boundary state condition 
\begin{align}\label{LCondition}
   \Big( L^{(1)}_{n} + L^{(2)}_{n} - \bar{L}^{(1)}_{-n} - \bar{L}^{(2)}_{-n} \Big) | B\rangle\rangle = 0
\end{align}
We deploy a general $O(2, \mathbb{C})$ reflection ansatz to connect the left- and right-moving oscillator modes across the two sectors:
\begin{align}\label{gluing condition}
    a^{(1)}_n &= -\cos 2\theta \; \bar{a}_{-n}^{(1)} - \sin 2\theta\; \bar{a}_{-n}^{(2)} \nonumber\\
    a^{(2)}_n &= -\sin 2\theta \; \bar{a}_{-n}^{(1)} + \cos 2\theta \; \bar{a}_{-n}^{(2)} 
\end{align}
The above then yields the matching equations 
\begin{align}
   - Q^{(1)} \cos 2\theta - Q^{(2)} \sin 2\theta + Q^{(1)} &= 0 \nonumber \\
   - Q^{(1)} \sin 2\theta + Q^{(2)} \cos 2\theta + Q^{(2)} &= 0
\end{align}
We can then solve for the gluing angle $\theta$:
\begin{align}\label{theta_complex_dilaton}
    \tan \theta = \frac{Q^{(2)}}{Q^{(1)}} = \frac{q_R - i q_I}{q_R + i q_I}
\end{align}
By writing the complex background charges in polar coordinates we can obtain a different parameterization for the gluing angle: $Q^{(1)} = r e^{i\varphi}$ and $Q^{(2)} = r e^{-i\varphi}$ (where $\tan \varphi = q_I / q_R$). Then the tangent of the gluing angle reduces directly to a pure phase,  $\tan \theta = e^{-2i\varphi}$. In this form, $\theta$ is given by 
\begin{align}\label{gluing angle}
    \theta = \frac{\pi}{4} - \frac{i}{2} \ln\left( \tan\left(\frac{\pi}{4} + \varphi\right) \right)
\end{align}
Because $\theta$ contains an explicit imaginary component proportional to the non-Hermitian charge ratio $\varphi$, the gluing condition defines a pseudo-orthogonal $SO(2, \mathbb{C})$ rotation in the field space. Physically, this means the interface actively mixes real and imaginary components of the field gradients.

\subsubsection{Boundary State and interface entropy}
The boundary state can be written as
\begin{align}
    | B \rangle\rangle = \mathcal{N} \exp\Big( \sum_{n>0} \frac{1}{n} S_{ij} a_{-n}^{(i)}\bar{a}_{-n}^{(j)} \Big) | B\rangle_0 
\end{align}
with S being the gluing matrix 
\begin{align}
    S = \begin{pmatrix}
        -\cos 2\theta    &   -\sin 2\theta \\
       -\sin 2\theta   & \cos 2\theta  
    \end{pmatrix}
\end{align}
$\ket{B}_0$ is determined by the gluing condition \ref{LCondition}, and as shown in appendix \ref{appendixDilaton}, the full boundary state is given by 
\begin{align}
\mid B\rangle \rangle&= {\cal N} \exp\Big( \sum_{n>0} {1\over n} S_{ij} a_{-n}^{(i)} \bar a_{-n}^{(j)}\Big)\int dp e^{ i p y^{\perp}  }\ket{ -i Q^{(1)}  + \sqrt{2}{Q^{(2)}\over \hat Q} p} \otimes   \ket{ -i Q^{(2)}  - \sqrt{2}{Q^{(1)}\over \hat Q}p } 
\end{align}
With normalization $\mathcal{N}$. Here, the integration is done over the Dirichlet momentum $p$, with $y^{\perp}$ being the Dirichlet position of the interface. The shift in the zero mode momentum caused by the background charge is offset by the shift in the central charge when evolving the state along the cylinder. In appendix \ref{gFactorApp}, we derive this explicitly via the Cardy  condition on the cylinder and determine the interface entropy which is expressed in terms of the complex gluing angle $\theta$
\begin{align}
    g= {\cal N} ={1\over \sqrt{\sin 2\theta}}
\end{align}
We note that this is the same as the result in the Janus case (\ref{janusgfactor}).

\subsubsection{Reflection and transmission}
We show in appendix \ref{lindilRT}, that a slightly  modified version of the formula for the reflection and transmission coefficients found in \cite{Quella:2006de} can be used to calculate the reflection and transmission coefficients. Interestingly the result has the same form as for the Janus case

\begin{equation}
    \begin{split}
        \mathcal{T} = \sin^2(2\theta),\quad \mathcal{R} = \cos^2 (2\theta)\\
    \end{split}
\end{equation}
The gluing angle $\theta$ is 
complex  but the  coefficients still satisfy  a conservation condition $\mathcal{T} + \mathcal{R} = 1$. Substituting in (\ref{gluing angle}), we then find 
\begin{equation}
    \mathcal{T} = \sec^2(2\varphi) \geq 1,\quad \mathcal{R} = -\tan^2(2\varphi) \leq 0
\end{equation}
Comparing with \ref{janusT}, we see that the coefficients for both the linear dilaton and the Janus case have similar qualitative features, with $\mathcal{T} \geq 1$ for both.

\section{Summary and discussion}

In this paper we have investigated complex CFTs from a holographic and field theoretic point of view. 
Such theories were first discussed in the context of walking RG-flows where two real fixed points merge and move into to complex conjugate points in the complex coupling plane.
We presented a holographic proof of the Im-flip property, which was previously argued for in \cite{Gorbenko:2018ncu} using intuitive arguments.   

In the holographic setting we constructed complex CFTs using a model of scalars coupled to gravity. For massless scalars one can give the scalar a complex expectation value. For scalars with a potential the potential can be chosen such that an extrema is at a complex value. In the former case some correlation functions will be complex while the central charge and scaling dimensions will stay real whereas in the latter case the holographic central charge and scaling dimensions of operators dual to scalar fluctuations will be complex.

We are particularly interested in cases where there are two CFTs which are complex conjugate to each other and there exists conformal interfaces between such theories.  For massless scalars such interfaces correspond to Janus solutions with an imaginary deformation parameter, which we investigated. The non-unitarity of the CFTs is reflected in  the fact that reflection and transmission coefficients while real do not satisfy unitarity bounds. We have shown that on the holographic side, the imaginary distance bound which was obtained for wormhole solutions in \cite{Maldacena:2026jqd,DiUbaldo:2026rly}  is satisfied for complex Janus solution.  We have also generalized this bound to the case of Janus solutions where the scalars are given by a nonlinear sigma model, in which the difference between scalars is replaced by a geodesic distance bound.  Note that for the single free boson the identification (\ref{janus_couplings}) the bound imaginary distance bound translates into the statement that the real part of the free boson coupling is positive such that euclidean path integral is convergent. It would be interesting to understand the general bound on the geodesic distance  we obtained on the holographic side in appendix \ref{app:c}, from a CFT point of view.

Furthermore we considered numerical solutions corresponding to RG-flow interfaces where the scalars approach  complex conjugate extrema on either side of the interface, in a simple model of a scalar with potential.  The numerical solutions were based on a model potential which is not realistic for all field values. In addition, small changes to initial conditions can  lead to drastically different solutions. This is to be expected of a nonlinear system with oscillating behavior and complex valued fields. It would be beneficial to find, if possible, examples of analytically solvable flows, specifically in the context of gauged supergravity. If such a case were to be found, holographic observables can also be computed without resorting to extensive numerical studies.

We have constructed some examples of conformal interfaces between conjugate CFTs on the field theory side. For the Janus case this can be obtained by considering the real Janus CFT and analytically continuing the deformation parameter. We constructed the boundary state, for which we computed the g-factor and the reflection and transmission coefficients. Further, we saw how the agreement to second order in the deformation parameter between the holographic and CFT sides is inherited from the real case.

For the Janus deformation the central charge and scaling dimension of operators stay real in order to construct an interface between complex CFTs with conjugate central charges and scaling dimensions we considered an interface the free noncompact boson with complex conjugate background charges on each side. The fact that the boson is still free allows us to construct the boundary state where the (complex) gluing angle is determined by the ratio of the background charges. The boundary state allows us to calculate g-factors and reflection and transmission amplitudes. A similar approach was used to construct open worldsheets \cite{Collier:2024mlg} for the complex Liouville string \cite{Collier:2024kmo,Collier:2024kwt}, it would be interesting to consider conformal interfaces for complex conjugate Liouville CFTs where we would not be limited to the case of a total charge adding up to $c=26$ in order to constitute a matter theory for the bosonic string.  It would be interesting to construct conformal interfaces for other complex conjugate CFTs such as the complex CFTs associated with the Potts model for $Q>4$ \cite{Gorbenko:2018dtm, Jacobsen:2026bvg}. It is possible that the results of the present paper can be useful since one approach to the Potts model is using the Coulomb-gas approach \cite{diFrancesco:1987ses,Dotsenko:1984nm} for general $Q$. It may also be interesting to use other approaches such as lattice formulations or loop models (see e.g. \cite{Jacobsen:2012zz}) to construct such interfaces. Many of the holographic constructions  and calculations can be generalized to higher dimensions. We leave these interesting questions for future work.

\acknowledgments
M.G. is grateful to the Centro de Ciencias de Benasque Pedro Pascual and the program "Gauge theory, supergravity and superstrings" for hospitality during completion of this work.
Both authors  are grateful to the Bhaumik Institute for support.

\newpage

\appendix

\section{CFT conventions}
\label{app:a}
In this appendix we describe our conventions for the free boson CFT with and without background charge for the convenience of the reader.

\subsection{Free noncompact boson}
\label{app:a1}
The  action is given by 
\begin{equation}
    S  = \frac{1}{8\pi}\int d^2 \sigma \sqrt{g } \, \partial_\alpha  X \partial^\alpha  X\\
\end{equation}
 If we take flat metric $g_{ab}=\delta_{ab}$ and go to complex coordinates with conventions
\begin{align}
    \int d^2 \sigma \sqrt{g} = {1\over 2} \int dz\bar dz , \quad \partial_a X \partial^{a}X= 2 \partial X  \bar \partial    X
\end{align}
The action becomes
\begin{align}
    S= {1\over 2\pi} \int d^2z  \partial  X  \bar \partial   X
\end{align}
It follows that the holomorphic and antiholomorphic components of the stress tensor 
\begin{align}\label{stresstenNonCom}
    T(z)  &= - :\partial X \partial X : , \quad \quad T(\bar z)  = - :\bar \partial X \bar \partial X : 
\end{align}
OPEs and correlation functions can be evaluated using the contractions
\begin{align}\label{opefree}
    \langle X(z_1,\bar z_1)   X(z_2,\bar z_2) \rangle &= -{1\over 2} \ln |z_1-z_2|^2 , \quad \quad 
    \langle  \partial X(z_1)   \partial X(z_2)\rangle = -{1\over 2 } {1\over (z_1-z_2)^2}
\end{align}
The mode expansion of $\partial X$ is given by
\begin{align}\label{modedX}
    \partial X &=- i  {1\over \sqrt{2} }\sum_n a_n z^{-n-1}, \quad\quad
     \bar \partial X =- i {1\over \sqrt{2} } \sum_n \bar a_n \bar z^{-n-1}
\end{align}
The OPE  (\ref{opefree})  leads  to the commutation relations
\begin{align}\label{comrel}
    [a_m,a_n]=m \delta_{m+n,0}, \quad   [\bar a_m,\bar a_n]=m \delta_{m+n,0}
\end{align}
The Virasoro generators are given by 
\begin{equation}
    L_m  = \frac{1}{2}\sum_{n\in \mathbb{Z}}:a_{m-n}a_n:\\
\end{equation}
And the central charge  of a single free boson is $c= 1$.  A boson with a noncompact target space does not have any winding modes and the momentum is equal for left and right moving modes.

\subsection{Linear dilaton theory}
\label{app:a2}
The linear dilaton theory is defined by adding a term proportional to $R^{(2)}[g]$  to  the worldsheet to the action of the free boson defined above.
\begin{equation}\label{lindilac}
    S  = \frac{k}{4\pi}\int d^2 \sigma \sqrt{g } \, \partial_\alpha  X \partial^\alpha  X+ {1\over 4\pi}  \int d^2 \sigma \sqrt{g} Q X R^{(2)}[g]\\
\end{equation}
Here $k$ is called the stiffness and $Q$ the background charge. $k$ can be set to one by rescaling the bosonic field, and we follow this convention.  On a flat worldsheet the second term  (\ref{lindilac}) and consequently  the OPE (\ref{opefree}), mode expansion (\ref{modedX}) and commutation relations (\ref{comrel}) are unchanged on the plane. However, the background charge term leads to a modification of the stress tensor due to the background charge
and with the stress energy tensor
\begin{align}\label{stressten}
    T(z)  &= - :\partial X \partial X : + Q \partial^2X, \quad \quad
     \bar T(\bar z)  = - :\bar \partial X \bar \partial X : +  Q \bar \partial^2X 
\end{align}
The stress tensor OPE is modified due to the presence of the background charge term
\begin{align}
    T(z)T(w)&= \frac{1}{2}\frac{1 + 6 Q^2}{(z-w)^4} +\frac{2}{(z-w)^2}\left(- :(\partial X(w))^2: +  Q  \partial^2X (w)
    \right)\nonumber \\
    &\quad + \frac{1}{z-w}\partial_w\left(-:(\partial X(w))^2: + Q \partial^2X (w)
    \right)+\cdots
\end{align}
Hence the central charge is shifted from $c=1$ 
\begin{equation}
    c = 1 +6  Q^2
\end{equation}
The conformal weight of the primary field $:e^{ikX}:$ will be shifted due  $h=\frac{k^2}{4 }+ i Q\frac{k}{2}$, due to the $T$ OPE 
\begin{equation}
    T(z):e^{ikX(w)}: = \left(\frac{k^2}{4}+iQ\frac{k }{2}
    \right)\frac{:e^{ikX(w)}:}{(z-w)^2} + \frac{\partial_w:e^{ikX(w)}:}{z-w}\\
\end{equation}

The background charge leads to an anomaly for the charge associated with the shift of the scalar field, such that an inner product  on the sphere is:
\begin{align}\label{spherean}
\langle p\mid p' \rangle = \delta(p-p'+ 2 i Q)
\end{align}
or alternatively as a shift of the charge under hermitian conjugation
\begin{align}
    a_0^\dagger =  a_0 -i{\sqrt{2}Q }, \quad  \bar a_0^\dagger =  \bar a_0 -i {\sqrt{2}Q },
\end{align}
whereas for the other modes we have the standard conjugation relations $a_n^\dagger = a_{-n}, \; \bar a_n^\dagger = \bar a_{-n},$.

\section{Boundary states for linear dilaton interfaces}\label{appendixDilaton}
After folding a conformal interface corresponds to a boundary state in the tensor product $CFT_1 \otimes \overline{CFT}_2$, which describes a boundary CFT in the tensor product.
\begin{align}
   \Big( L^{(1)}_{n}+ L^{(2)}_{n}-\bar L^{(1)}_{-n} -\bar L^{(2)}_{-n}\Big) \mid B\rangle \rangle&=0
\end{align}
The interface is between two linear dilaton theories with background charge $Q^{(1,2)}$  respectively.  The mode expansion of the stress tensor (\ref{stressten}) is given by
\begin{align}\label{mode-expld}
    L_{m}^{(i)} &= \frac{1}{2}\sum_n a_{m-n}^{(i)}a_n^{(i)} + \frac{iQ^{(i)}}{\sqrt{2}}a_{m}^{(i)}(m+1)
\end{align}\label{o2gluing}
The terms quadratic in the modes $a_n$ are glued using a $O(2)$ matrix.
\begin{align}
  a_{n}^{(i)} = S_{ij} \bar a_{-n}^{(i)}, \quad \quad   S = \begin{pmatrix}
        -\cos 2\theta    &   -\sin 2\theta \\
       -\sin 2\theta   & \cos 2\theta  
    \end{pmatrix}
\end{align}
Note that for a boson without a background charge this gluing condition corresponds to a D1 brane on an angle \cite{Bachas:2007td}. The T-dual condition where the gluing is a proper $SO(2)$ matrix is not consistent with the linear gluing conditions which are new to the case with background charge.  For the term linear in the oscillator modes  one uses the gluing condition (\ref{mode-expld}), one has to distinguish terms which are proportional to $m$ and terms which are not. The former relate the rotation angle $\theta$ to the background charges of the interface.
\begin{align}
   -Q^{(1)}\cos 2\theta - Q^{(2)}\sin 2\theta + Q^{(1)} &=0\nonumber \\
        -Q^{(1)}\sin 2\theta + Q^{(2)}\cos 2\theta + Q^{(2)} &=0
\end{align}
which can be solved for:
\begin{align}\label{rotangle}
  \tan\theta ={Q^{(2)}\over Q^{(1)}} 
  \end{align}
  The $m$ independent terms  also contain terms from the quadratic piece due to the presence of $a_m a_0$ pieces in (\ref{mode-expld}), yielding the  following conditions for the zero modes $a_0^{(1,2)} $\footnote{Recall that we consider the noncompact bosons  without  winding modes and $a_0^{(i)}=\bar a_0^{(i)}$.}
   \begin{align}
        -a_0^{(1)}(1+\cos 2\theta) - a_0^{(2)}\sin 2\theta - i\sqrt{2}Q^{(1)} &= 0 \nonumber \\
        a_0^{(2)}(\cos 2\theta - 1)-  a_0^{(1)}\sin 2 \theta - i\sqrt{2}Q^{(2)} &= 0 
\end{align}
The possible values for the zero modes which solve these equations are
\begin{align}\label{zeromglue}
        a_0^{(1)} &= -\frac{i}{\sqrt{2}}Q^{(1)} + \frac{Q^{(2)}p}{\hat Q}\nonumber \\
        a_0^{(2)} &= -\frac{i}{\sqrt{2}}Q^{(2)} - \frac{Q^{(1)}p}{\hat Q}
\end{align}
where we have defined $\hat Q = {\sqrt{(Q^{(1)})^2 + (Q^{(2)})^2}}$ and 
 $p$ is an arbitrary real number. The first term is a fixed background momentum which is half of the one which saturates the sphere anomaly  (\ref{spherean}). This is due to the fact that a boundary states creates a hole in a  worldsheet making the sphere into a disk. Hence the  charge anomaly is halved since the Euler character of the disk is half of the one of the sphere. The second term in (\ref{zeromglue}) corresponds to a momentum associated with Dirichlet boundary conditions for the rotated coordinate $X^\perp= {\small {1\over \hat Q}( Q^{(2)} X^1- Q^{(1)} X^2)}$.  The complete boundary state is then given by
\begin{align}
\mid B\rangle \rangle&= {\cal N} \exp\Big( \sum_{n>0} {1\over n} S_{ij} a_{-n}^{(i)} \bar a_{-n}^{(j)}\Big)\int dp \; e^{ i p y^{\perp}  }\ket{ -i Q^{(1)}  + \sqrt{2}{Q^{(2)}\over \hat Q} p } \otimes   \ket{ -i Q^{(2)}  -\sqrt{2} {Q^{(1)}\over \hat Q}p  } 
\end{align}

\subsection{Interface $g$-factor}\label{gFactorApp}
For a boundary CFT the $g$ factor, which from the point of view of D-branes is related to the brane tension, is given by the overlap of the boundary state with the vacuum (\cite{Quella:2006de}, \cite{Bachas_2002}). Note that due to the background charge this overlap vanishes  because of (\ref{spherean}).      One can replace the vacuum with the simplest state with nonzero overlap,  and obtains
\begin{align}
 g&= \langle  i Q^{(1)} \mid \otimes \langle i Q^{(2)} \mid B \rangle\rangle = {\cal N}
\end{align}

In order to obtain an expression for $\mathcal{N}$, we must satisfy the Cardy–Lewellen crossing condition for the boundary state on a cylinder. Thus, we equate the cylinder amplitude of the interface in the open string channel to that in the closed string channel. On the cylinder, the mode expansion for $L_0$ is given by 
\begin{equation}
    \begin{split}
        L_0^{(i)} = \frac{1}{2}(a_0^{(i)})^2 + \sum_{n>0} a_{-n}^{(i)}a_n^{(i)} + \frac{iQ^{(i)}}{\sqrt{2}}a_0^{(i)} \\
    \end{split}
\end{equation}
With the closed string Hamiltonian on the cylinder being 
\begin{equation}
    H_{cl} = L_0^{(1)} + L_0^{(2)} + \bar{L}_0^{(1)} + \bar{L}_0^{(2)} - \frac{c_{total}}{12}\\
\end{equation}
We can then compute the explicit form of $H_{cl}$ using \ref{zeromglue}, and we note that the background charge completely cancels out:
\begin{equation}
   H_{cl} =  p^2-\frac{1}{6}\\
\end{equation}
With the amplitude being independent of the background charge, we can calculate the open cylinder amplitude in the compact boson case and take the noncompact limit. As discussed in \cite{Quella:2006de} and \cite{PhysRevLett.67.161}, finding the amplitudes is equivalent to placing the theory on a torus of length $\ell = \sqrt{\ell_1^2 +\ell_2^2}$ and volume $V = \ell_1\ell_2$ the normalization can be found. Taking the noncompact limit is equivalent to taking $\ell_i = Q^{(i)}$, the normalization is given by \cite{Quella:2006de}
\begin{equation}
    \mathcal{N} = \frac{\ell}{\sqrt{2V}} = \sqrt{\frac{\hat{Q}^2}{2Q^{(1)}Q^{(2)}}} = \frac{1}{\sqrt{\sin 2\theta}}\\
\end{equation}
which we note has the same form as the normalization for the Janus case (\ref{janusgfactor}).

\subsection{Reflection and transmission coefficients}  \label{lindilRT}
The Reflection and transmission coefficients can be calculated using the expressions derived by Quella, Runkel and Watts \cite{Quella:2006de}. 
The reflection and transmission coefficients can be obtained from the following overlaps
\begin{align}\label{qrw}
    R_{ij} ={\langle 0\mid L_2^{(i)} \bar L_2^{(j)} \mid B\rangle\rangle\over \langle 0 \mid B\rangle \rangle}
\end{align}
with
\begin{align}
    {\cal R}&= {2\over c_1+c_2}\big(R_{11}+R_{22}\big), \quad \quad 
    {\cal T} = {2\over c_1 +c_2} \big(R_{12}+R_{21}\big)
\end{align}
However for the same reason as alluded to in the previous subsection this prescription has to be modified since the overlap with $\langle 0\mid$ in (\ref{qrw}) vanishes. Instead we propose to replace the vacuum with a primary field $\langle \phi \mid$ in $CFT^{(1)}\otimes \overline{CFT}^{(2)}$.
\begin{align}\label{qrw2}
    R_{ij} ={\langle \phi \mid L_2^{(i)} \bar L_2^{(j)} \mid B\rangle\rangle\over \langle \phi \mid B\rangle \rangle}
    \end{align}
    and 
    \begin{align}\label{newRT}
    {\cal R}&= {1 \over 4 h_\phi  +{c_1+c_2\over 2}}\big(R_{11}+R_{22}\big), \quad \quad 
    {\cal T} = {1 \over 4 h_\phi  +{c_1+c_2\over 2}}\big(R_{12}+R_{21}\big)
\end{align}
Note that this formula goes over to the original one in the case of vanishing background charge and $\langle \phi \mid =\langle 0\mid$. Furthermore the sum rule
\begin{align}
\sum_{ij} R_{ij} =  {\langle \phi  [L^{tot}_{2}, L^{tot}_{-2}] \mid B\rangle\rangle\over \langle  \phi \mid B\rangle \rangle} = 4 h_\phi + {c_1+c_2\over 2}
\end{align}
Implies that  with the normalization above we have conservation ${\cal R}+{\cal T}=1$. The particular primary we choose is 
\begin{equation}\label{phichoice}
    \langle \phi\mid   = {1\over \small{\sqrt{2}}} \Big( \langle iQ^{(1)}  +\sqrt{2} {Q^{(2)}\over \hat Q}p\mid \otimes\langle iQ^{(2)}  - {\sqrt{2}}{Q^{(1)}\over \hat Q}p\mid +\langle iQ^{(1)}  -\sqrt{2} 
    {Q^{(2)}\over \hat Q}p\mid \otimes\langle iQ^{(2)}  + {\sqrt{2}}{Q^{(1)}\over \hat Q}P\mid\Big)
\end{equation}
The $L$ operators are then given by

\begin{align}\label{newStressmodes}
        L_2^{(1)} &= \frac{1}{2}a_1^{(1)}a_1^{(1)} + \left(\frac{Q^{(2)}}{\hat Q} p  + i\sqrt{2}Q^{(1)}
        \right)a_2^{(1)}\nonumber\\
        L_2^{(2)} &= \frac{1}{2}a_1^{(2)}a_1^{(2)} + \left(-\frac{Q^{(1)}}{\hat Q }p  + i\sqrt{2}Q^{(2)}
        \right)a_2^{(2)}
\end{align}
With the above, it is easily checked that 
\begin{align}
      R^{11} + R^{22} &= \cos^2 2\theta + 2\cos 2\theta\left((Q^{(1)})^2 - (Q^{(2)})^{2})(2+ \frac{p^2}{\hat Q^2}
        \right)\nonumber\\ 
        &\quad + 4\sqrt{2}i \frac{Q^{(1)}Q^{(2)}}{\hat Q }p\cos 2\theta-4\sqrt{2}i \frac{Q^{(1)}Q^{(2)}}{\hat Q}p\cos 2\theta \nonumber\\
      R^{12} + R^{21} &= \sin^2 2\theta + 4\sin 2\theta\left(Q^{(1)}Q^{(2)}(2 + \frac{p^2}{\hat Q^2})
        \right) \nonumber \\
        &\quad -4\sqrt{2}i\sin 2\theta \frac{(Q^{(2)})^2 - (Q^{(1)})^2}{\hat Q}p+4\sqrt{2}i\sin 2\theta \frac{(Q^{(2)})^2 - (Q^{(1)})^2}{\hat Q}p
    \end{align}
Note that the terms linear in $p$ cancel due to the choice of $\bra{\phi}$ in (\ref{phichoice}) which is symmetric under $p\to -p$. Using (\ref{newRT}) along with (\ref{rotangle}) it is straightforward to show that 

\begin{align}
        \mathcal{R} & = \cos^2 (2\theta), \quad \quad  \mathcal{T} = \sin^2 (2\theta)
\end{align}
Note that the result is independent of the choice of momentum $p$ which lends credence to consider this result a universal formula for reflection and transmission coefficients. It is remarkable that the result has the same form as for the Janus  interface without background charges.\

\section{Complex Janus with Sigma model scalars}
\label{app:c}

In this appendix we consider complex Janus solutions for scalars which are defined by a nonlinear sigma-model metric which is given by action (\ref{actiona}) with a constant potential $V=-2$.  
Since massless fields are associated with marginal operators  on the CFT side, this manifold can be viewed as the holographic realization of the Zamolodchikov metric \cite{Zamolodchikov:1986gt} on the conformal manifold of the CFT.

Note that the scalar equation of motion  (\ref{secondorder}) for this case is  repeated  here for the reader's convenience.
\begin{align}\label{geoflow}
{d^2\phi^a\over du^2} +2 {dA\over du} {d \phi^a\over du} + \Gamma^{a}_{\; bc} (\phi) {d\phi^b\over du} {d\phi^c\over du} &=0 
\end{align}
has the form of a geodesic equation for the motion of the scalar on the moduli space with a friction term, which means that the  scalar  during the flow moves on a geodesic that is not affinely parameterized.  It follows from the geodesic equation that the length of the tangent vector on the scalar manifold satisfies the  following relation
\begin{align}
G_{ab} {d \phi^a\over du}  {d \phi^b\over du}  = \alpha ^2 \; e^{-4 A(u)} 
\end{align}
Plugging this expression into the last equation of (\ref{secondorder}) 
it follows that the gravitational equations of motion are exactly the same as for the single massless scalar, and the solution for $A(u)$ is given by (\ref{janusfun}) which depends only on the  integration constant  $\alpha$.
Hence the total geodesic length can be calculated by integrating
\begin{align}\label{complexdistance}
   \Delta s^2= \int_{-\infty}^{+\infty} du \; G_{ab} {d \phi^a\over du}  {d \phi^b\over du} &= \int_{-\infty}^{+\infty} du  {4 \alpha^2 \over \big(1+\sqrt{1-\alpha^2} \cosh(2u) \big)^2} \nonumber \\
    &= -4 -{8\over \alpha} \tanh^{-1}\left( {-1+ \sqrt{1-\alpha ^2}\over \alpha }\right)
\end{align}

For imaginary flows\footnote{There are    conditions on the real moduli space metric  $G_{ab}$ for  an imaginary flow to always be imaginary. This is due to the quadratic in $d\phi^a/du$ in (\ref{geoflow}). For example $G_{ab}(i \psi) = \pm G_{ab}(\pi)$ is sufficient.  } we continue the parameter  $\alpha \to i \gamma$ and (\ref{complexdistance}) is equal to $\sigma$  defined in (\ref{sigmares}) which determines the reflection and transmission coefficient.   Note that for imaginary $\alpha=i\gamma$ the geodesic distance on the moduli space is negative and $\gamma$  takes values between zero and infinity. The total geodesic length for the imaginary flow  is bounded by  
\begin{align}
    -4< \Delta s^2< 0
    \end{align}
 which generalizes the imaginary distance bound for a single scalar (\ref{imdistance}) to the case of a curved moduli space.

\newpage
\providecommand{\href}[2]{#2}\begingroup\raggedright\endgroup

\end{document}